\documentclass{aa}
\usepackage{graphicx}
\usepackage{natbib}
\usepackage{jour}
\bibpunct{(}{)}{;}{a}{}{,}

\begin{document}

\title{Cosmological Hydrodynamics with Adaptive Mesh Refinement}
\subtitle{A New High Resolution Code Called RAMSES}

\author{Romain Teyssier}
\institute{
Commissariat \`a   l'Energie Atomique, Direction  des  Sciences  de la
Mati\`ere, Service   d'Astrophysique, \\  
Centre d'Etudes de Saclay, L'orme des Merisiers, F-91191 Gif-sur-Yvette
Cedex\\
\email{Romain.Teyssier@cea.fr}
\and 
Numerical Investigations in Cosmology (NIC group), CEA Saclay.
}

\abstract{
A new N-body and hydrodynamical code,  called RAMSES, is presented. It
has been designed  to study structure formation  in the universe  with
high  spatial  resolution.   The   code  is  based on   Adaptive  Mesh
Refinement (AMR)  technique, with a tree  based data structure allowing
recursive grid refinements on a cell-by-cell basis.  The N-body solver
is   very    similar  to  the   one  developed   for  the   ART   code
\citep{kravtsov97},     with   minor    differences     in  the  exact
implementation.  The hydrodynamical solver is  based on a second-order
Godunov  method, a modern   shock-capturing  scheme known  to  compute
accurately the thermal history of  the fluid component.  The  accuracy
of the code is carefully estimated using various test cases, from pure
gas dynamical tests  to  cosmological ones.  The  specific  refinement
strategy used in  cosmological simulations is described, and potential
spurious  effects   associated  to  shock waves    propagation  in the
resulting AMR grid are discussed and found  to be negligible.  Results
obtained in a large  N-body and hydrodynamical simulation of structure
formation in a low density $\Lambda$CDM universe are finally reported,
with  $256^3$ particles and  $4.1\times  10^7$ cells  in the AMR grid,
reaching a  formal resolution of $8192^3$.   A convergence analysis of
different quantities, such as dark matter  density power spectrum, gas
pressure  power spectrum  and  individual haloes temperature  profiles,
shows that  numerical   results are  converging   down  to the  actual
resolution  limit of  the  code, and are   well  reproduced by  recent
analytical predictions in the framework of the halo model.
 }

\date{Accepted ??? / Received ???}

\offprints{Romain~Teyssier}
\mail{Romain.Teyssier@cea.fr}
\authorrunning{R. Teyssier}

\maketitle

\label{firstpage}
\section{Introduction}

Numerical  simulations of structure formation  in the universe are now
widely    used to  understand     the   highly non-linear   nature  of
gravitational  clustering. Dark matter is believed  to be the dominant
component in mass of the cosmological density field, with only a small
fraction (say 10~\%) in   baryons.  At  intermediate scales, such   as
galaxy clusters, dark  matter still dominates the total  gravitational
mass,  but  the introduction  of a   gaseous component appears   to be
unavoidable, since X-ray or Sunyaev-Zeldovich  observations of the hot
intracluster medium  give us  strong constraints  on the structure  of
galaxy clusters.  At smaller  scales,  gas cooling and fluid  dynamics
play a dominant role in  the structure of galaxy-size object. Although
baryons can be   described to first  order  as an  hydrostatic ionized
plasma  trapped in  dark   matter potential wells,   the complexity of
hydrodynamical  processes  such  as  shock  heating, atomic  radiation
cooling and, ultimately, star formation requires an accurate treatment
of the baryonic component.

For a  cosmological simulation to  be realistic, high mass and spatial
resolution  are needed.  While  the former is   related to the initial
number of       degrees  of   freedom  (usually     ``particles''   or
``wavelengths'')  in the computational  volume,  the latter is usually
related to the  numerical   method specifically used to   compute  the
particles trajectory. For a $\Omega=1$  universe, using a sufficiently
large volume of, say, 100 Mpc h$^{-1}$ aside, we need at least $256^3$
particles  to describe  $L_{\star}$  galaxies with  100  particles. In
order to resolve  the internal radial structure of  such haloes with at
least  10 resolution elements, we need  a spatial resolution of 10 kpc
h$^{-1}$ or equivalently a dynamical range of $10^4$.

The  Particle-Mesh  method \citep{hockney81, klypin83}  is perhaps the
simplest  and   fastest N-body   algorithm  for solving  gravitational
dynamics, but it is limited by computer resources to a dynamical range
of $10^3$. The P3M method \citep{hockney81, edfw85} can reach a higher
spatial resolution, by adding a small scale component to the PM force,
directly computed from the two-body interactions ("Particle-Particle")
between neighboring particles.    This method suffers however from   a
dramatic  increase  in CPU time as   clustering develops and  as short
range forces  become  dominant.  An  improvement  of this   method was
therefore developed for the  AP3M code \citep{couchman91}: a hierarchy
of recursively  refined    rectangular grids is   placed in  clustered
regions where  a local  PM  solver is activated   to  speed up  the PP
calculations.  Another N-body method which  can achieve high dynamical
range is the TREE code \citep{barnes86,bouchet88}, which properly sort
neighboring particles in   a recursive  tree  structure.  Long   range
interactions are computed using multipole expansion and low resolution
nodes of the tree, while short range interactions are computed using a
PP approach between particles belonging to the same leaf of the tree.

The idea  of using Adaptive Mesh Refinement  (AMR)  for N-body methods
appears as a natural generalization of both AP3M and TREE codes, since
they  use  a hierarchy   of nested   grids   to increase  the  spatial
resolution  locally.      The   recently      developed    ART    code
\citep{kravtsov97} offers, in  this respect,  the first implementation
of a grid-based high-resolution N-body code, where the mesh is defined
on a recursively refined  spatial tree.  ART  takes advantage of  both
the speed of a mesh-based Poisson solvers and the high-dynamical range
and flexibility obtained with a tree structure.   Since no PP force is
considered in    the ART method, the   resolution  is not  uniform (as
opposed to AP3M  and TREE codes), but proportional  to  the local cell
size of the grid.  The  grid is continuously  refined or de-refined in
the course of   the simulation,  to ensure  that  the  mean number  of
particles  par  cell remains  roughly    constant (around 10).    This
``quasi-Lagrangian'' approach ensures that two-body relaxation remains
unimportant \citep{knebe00}.

The gaseous component, baryons, can be described  using one of several
hydrodynamical  methods widely  used today in   cosmology. They can be
divided  into three groups:  Lagrangian schemes,  Eulerian schemes and
intermediate schemes.

1-  Lagrangian  schemes or  quasi-Lagrangian  schemes \citep{gnedin95,
pen95} are based on a moving mesh that closely follows the geometry of
the flow for a constant number of grid points.  The grid adapts itself
to collapsing fluid elements, but suffers  from severe mesh distortion
in    the   non   linear      stage  of    gravitational    clustering
\citep{gnedin96}. The coupling with one of the aformentioned N-body
solvers is also non-trivial (ibid).

2- Eulerian schemes are usually based on uniform Cartesian mesh, which
make them suitable for a coupling with the traditional, low-resolution
PM solver  \citep{cen92, ryu93,   teyssier98, chieze98}.  They  suffer
however from limited dynamical range.

3-   Smooth Particles   Hydrodynamics (SPH) can     be thought  as  an
intermediate solution. This is a  particle-based method, which follows
the Lagrangian evolution of the flow, but in which resolution elements
are  defined as appropriate  averages   over 50 neighboring  particles
\citep{gingold77,evrard88,  hernquist89}.  This ``smoothing   kernel''
defines the effective Eulerian   resolution of  the method.  The   SPH
method  offers also the  possibility of a  straightforward coupling to
particle-based  N-body solver like the AP3M   or TREE codes.  The main
drawback of  the   SPH  method  is  its  relative poor   discontinuity
capturing capabilities (one needs at least 50 particles per resolution
element in order to properly describe sharp features, like shock waves
or contact    surfaces) and  the  fact  that  it relies  on  the
artificial viscosity method to capture shock waves.

One of the  most promising hydrodynamical  method at this  time is the
AMR   scheme,  described     originally   in    \citet{berger84}   and
\citet{berger89}. The original AMR method is an Eulerian hydrodynamics
scheme,  with a hierarchy   of  nested grids covering  high-resolution
regions of the  flow.  The building blocks  of the computational  grid
are therefore rectangular  patches of  various sizes, whose  positions
and aspects ratio are optimized  with respect to flow geometry,  speed
and  memory     constraints.  Let's  call    this   spatial  structure
``patch-based AMR''.  An   alternative method was proposed  by several
authors  \citep[see][  and reference therein]{khokhlov98} where parent
cells are  refined into children cells, on  a cell-by-cell  basis.  As
opposed to the original patch-based AMR, let  us call this last method
``tree-based AMR'', since  the  natural data structure  associated  to
this scheme is a recursive tree structure.  The resulting grid follows
complex flow geometry more closely, at  the price of a data management
which is more complicated  than patch-based AMR.  These  two different
adaptive mesh structures can    be  coupled to any grid-based    fluid
dynamics scheme.  It  is worth mentioning that modern  high-resolution
shock capturing   methods  are   all grid-based and    have  number of
interesting features:  they  are stable  up to  large Courant numbers,
they are striclty conservative for  the Euler equations and they
are able  to  capture discontinuities  within only  few cells.   Among
several schemes,  higher order  Godunov   methods appear to  be   more
accurate and to be easy to generalize in 3 dimensions.

The original patch-based AMR, based  on the Piecewise Parabolic Method
(PPM: a third-order Godunov scheme), was recently adapted to cosmology
\citep{bryan97,abel00}.   The hydrodynamical scheme  was coupled to
the AP3M N-body  solver (without  using  the PP interaction   module).
This  choice  is natural  since  both codes use   a set of rectangular
patches to cover high-resolution regions of the flow.

In   this paper,   an  alternative solution   is  explored: coupling a
tree-based AMR hydrodynamical scheme   to the N-body  solver developed
for the ART code  \citep{kravtsov97}.  This solution seems indeed more
suitable  to the hierarchical  clustering picture where a very complex
geometry  builds up,  with a   large  number of small  clumps  merging
progressively   to  form large virialized  haloes   and filaments.  The
number of  grids required to  cover efficiently  all these small haloes
is so large, that it renders a patch-based approach less efficient.

In this  paper,  a  newly  developed N-body  and  hydrodynamical code,
called RAMSES, is presented.  It is a tree-based AMR using the ``Fully
Threaded Tree''  data structure   of \citet{khokhlov98}.  The   N-body
solver is largely inspired   by the ART code  \citep{kravtsov97}, with
some  differences  in the  final  implementation.  The  hydrodynamical
solver is a second-order Godunov scheme for perfect gases (also called
Piecewise Linear Method or PLM).  In Section~\ref{sec:nummethods}, the
N-body and hydrodynamical algorithms  developed for RAMSES are briefly
described, with emphasis put  on the original solutions discovered  in
the course of this work.  In Section~\ref{sec:tests}, results obtained
by  RAMSES for standard test  cases  are presented, demonstrating  the
accuracy of the method.  Pure  hydrodynamical problems are  considered
first, showing  that shocks and  contact surfaces are well captured by
the tree-based AMR scheme.

Great care is  taken to demonstrate  that refining  the mesh in  shock
fronts  can be avoided  in cosmological contexts.  Indeed, potentially
spurious  effects associated  to the AMR  grid  remains low enough  to
apply the  method safely using   refinements  in high density  regions
only.  In Section~\ref{sec:simucosmo}, results of a large cosmological
simulation using $256^3$ particles,  with coupled gas and dark  matter
dynamics,    are  reported   and   compared   to  various   analytical
predictions. In Section~\ref{sec:conclusion}, the results presented in
this paper are summarized, and future projects are discussed.

\section{Numerical methods}
\label{sec:nummethods}

The modules  used in  RAMSES can  be divided  into  4 parts:  the  AMR
service routines, the   Particle  Mesh routines,  the Poisson   solver
routines and the hydrodynamical  routines.  The  dimensionality, noted
$\rm dim$, can be anything among 1, 2 or 3.

\subsection{Adaptive Mesh Refinement}

The fundamental data structure in RAMSES  is called a ``Fully Threaded
Tree'' (FTT) \citep{khokhlov98}.  Basic elements are not single cells,
but rather  groups of $2^{\rm dim}$  sibling cells called  {\em octs}.
Each oct  belongs  to a  given  level of refinement labeled  $\ell$. A
regular  Cartesian grid, called  the coarse grid,  defines the base of
the tree structure ($\ell=0$).  In order to access all octs of a given
level, octs are sorted  in a double linked   list.  Each oct at  level
$\ell$ points to  the previous and the  next  oct in the level  linked
list, but also to the parent cell at level $\ell-1$, to the $2
\times {\rm dim}$ neighboring  parent cells at  level $\ell-1$  and to
the $2^{\rm  dim}$  child octs at level  $\ell+1$.  If a  cell  has no
children,  it is called  a  \emph{leaf} cell, and  the pointer  to the
child oct is  set  to \emph{null}. Otherwise,  the  cell  is called  a
\emph{split  cell}.  In order to store  this particular tree structure
in memory, one needs therefore 17  integers per oct for ${\rm dim}=3$,
or equivalently $2.125$ integers per cell.

In   RAMSES, time integration can be   perfomed  in principle for each
level  independantly. Only  two   time stepping algorithms   have been
implemented  so far: a single  time step scheme   and an adaptive time
step scheme.  The single time  step algorithm consists in  integrating
the equations from   $t$ to $t+\Delta   t$, with  the  same  time step
$\Delta  t$ for all levels.  The  adaptive time step algorithm, on the
other hand, is similar to a ``W cycle''  in the multigrid terminology.
Each level is evolved in time with its own time  step, determined by a
level   dependant CFL stability   condition.  Consequenty,  when level
$\ell=0$ is  advanced  in time  using  one  coarse time   step,  level
$\ell=1$ is  advanced  in time  using two  time steps,  level $\ell=2$
using  4 time steps,  and  so on.  An  additional  constraint on these
level dependant time  steps comes from synchronization, namely $\Delta
t _{\ell} = \Delta t _{\ell+1} ^1 + \Delta t _{\ell+1} ^2$.

Within one time step and for each level, each operation is perfomed in
the following  way: a sub-sample of  octs  is first gathered  from the
tree.  Gathered cells can then be  modified very efficiently on vector
or parallel  architectures.  Finally, updated quantities are scattered
back to the tree.   When one needs to access  neighboring cells (in
order to compute  gradients  for example),  it is  straightforward  to
obtain the   neighboring  oct addresses from the   tree,   and then to
compute the neighboring cell addresses.

The two main routines used to  dynamically modify the AMR structure at
each time step are now described.

\subsubsection{Building the refinement map}
\label{sectrefinemap}

The  first step consists in marking  cells for refinement according to
\emph{user-defined} refinement criteria,  within the  constraint given
by a strict refinement rule: any  oct in the  tree structure has to be
surrounded  by $3^{\rm  dim}-1$  neighboring parent  cells.  Thanks to
this rule, a smooth transition in spatial resolution between levels is
enforced, even  in  the diagonal directions.    Practically, this step
consists in three  passes through each  level, starting from the finer
level $\ell_{\rm max}$ down to the coarse grid $\ell=0$.

\begin{enumerate}
\item If a split cell contains a children cell that is marked or
already refined, then mark it for refinement
\item Mark the $3^{\rm dim}-1$ neighboring cells.
\item If any cell satisfies the user-defined refinement criteria,
then mark it for refinement.
\end{enumerate}

One  key ingredient still missing in   this procedure is the so-called
``mesh smoothing''. Usually, refinement  are activated when  gradients
(or  second derivatives) in  the     flow variables exceed a     given
threshold.   The  resulting refinement  map  tends   to  be ``noisy'',
especially  in smooth  part  of  the flow  where gradients  fluctuates
around   the   threshold.   \citet{khokhlov98}    describes   a   very
sophisticated method based on a reaction-diffusion operator applied on
the refinement map.  I  prefer to use  here  the simpler   approach of
\citet{kravtsov97} in  the ART code, where a  cubic buffer is expanded
\emph{several times}  around marked  cells.  The number of   times one
applies the  smoothing operator on the refinement  map  is obviously a
free parameter.  This parameter is  noted $n_{\rm expand}$. In the ART
code, this operator is applied twice at each time step.  In RAMSES, it
is usually  applied   only once,  since, as   we see below,   boundary
conditions are defined for each level in a slightly more sophisticated
way  than in ART,  using \emph{buffer regions} (see \S~\ref{poisson}).
Therefore, the extra mesh  smoothing used in ART can  be thought as  a
way of creating the equivalent of the buffer regions in RAMSES.

Note that the exact method implemented here  and in the ART code leads
to a convex   structure for  the  resulting mesh,  that  is  likely to
increase the overall stability of  the algorithm. Note also that  only
refinement criteria  are necessary in RAMSES:  no \emph{de-refinement}
criteria  need  to be specified  by the  user.   This  is an important
difference   compared     to    other  approaches   \citep{kravtsov97,
khokhlov98}.

\subsubsection{Modifying the tree structure}

The next step  consists  in  splitting  or destroying children   cells
according to the refinement  map.  RAMSES performs two passes  through
each level,  starting from the coarse grid  $\ell=0$,  up to the finer
grid $\ell_{\rm max}$:

\begin{enumerate}
\item If a leaf cell is marked for refinement,
then create its child oct.
\item If a split cell is not marked for refinement,
then destroy its child oct.
\end{enumerate}

Creating or destroying a child oct is a  time-consuming step, since it
implies reorganizing the tree structure.  Thanks  to the double linked
list   associated to  the  FTT tree    structure,  this is done   very
efficiently by  first disconnecting the child  oct from the  list, and
then  reconnecting  the list  in between the   previous and next octs.
Note   however that this  operation can    not  be vectorized.  It  is
important to stress that this operation  is applied at each time step,
but \emph{for a very  small number of  octs}.  In other word,  at each
time step, the mesh  structure is not rebuilt from  scratch, but it is
slightly modified,   in  order to  follow the  evolution  of the flow.
Since  the refinement  map has been  carefully built  during  the last
step, the refinement rule should be satisfied by construction. This is
however not   the case  if one   uses  the adaptive  time step  method
described in Section~\ref{timestepping}.  In this  case, a final check
is performed before splitting  leaf cells.  If  the refinement rule is
about to be violated, leaf cells are not refined.

\subsection{N-body solver}

The N-body scheme used in RAMSES is similar in many aspects to the ART
code of \citet{kravtsov97}.  Since the ART code was not publicily
available at the time  this work was initiated, a  new code had  to be
implemented from scratch.  I briefly recall here the main ingredients
of     the method,   outlining   the differences      between  the two
implementations.

A collisionless  N-body  system  is  described by  the  Vlasov-Poisson
equations, which, in terms of particles (labeled by ``p''), reads

\begin{equation}
\label{nbody}
\frac{d{\bf x}_p}{dt} = {\bf v}_p~~{\rm and}~~\frac{d{\bf v}_p}{dt} = - \nabla_x \phi
~~{\rm where}~~ \triangle_x \phi = 4\pi G \rho
\end{equation}

Grid-based N-body  schemes, such   as the  standard  PM,  are  usually
decomposed in the following steps:

\begin{enumerate}
\item Compute the mass density $\rho$ on the mesh using a ``Cloud-In-Cell''
(CIC) interpolation scheme.
\item Solve for the potential $\phi$ on the mesh using the Poisson equation.
\item Compute the acceleration on the mesh using a standard finite-difference
approximation of the gradient.
\item Compute each particle acceleration using an inverse
CIC interpolation scheme.
\item Update each particle velocity according to its acceleration
\item Update each particle position according to its velocity.
\end{enumerate}

The  specific  constraints of a tree-based   AMR N-body solver are now
discussed in more details.

\subsubsection{The particle linked list}

Since we are dealing with an AMR grid, we need  to know which particle
is interacting with  a given cell.  This is done  thanks to a particle
linked list.  Particles belong to a given  oct, if their position fits
exactly into the oct boundaries.   All particles belonging to the same
oct are linked together. In  order to build this  linked list, we have
to store the position  of each  oct  in the tree  structure. Moreover,
each oct needs to have access to the address  of the first particle in
the  list and to  the  total   number of  particles contained  in  its
boundaries. We need therefore  to store these  two new integers in the
FTT tree structure.

The particle linked  list is built in a  way similar to the TREE code:
particles are first divided among the octs sitting at the coarse level
$\ell=0$.    Each individual linked  list  is then recursively divided
among the children octs, up  to the finer level $\ell=\ell_{\rm max}$.
Going from level $\ell$  to level $\ell+1$  implies removing from  the
linked  list particles sitting  within split  cell  boundaries.  Going
from level $\ell+1$ to level  $\ell$ implies reconnecting the children
linked lists  to the parent one.  In  the adaptive time step  case, in
order   to avoid rebuilding  the   whole tree from  the coarse  level,
particle positions are checked against parent  octs boundaries and, if
necessary, are passed to neighboring octs using the information stored
in the FTT tree.

\subsubsection{Computing the density field}

The density  field is  computed  using  the  CIC interpolation  scheme
\citep{hockney81}. For each level,  particles sitting inside level $\ell$ 
boundaries are first   considered.  This can be  done  using the level
$\ell$ particle linked  list.   Particles sitting outside  the current
level, but whose  clouds  intersect the corresponding volume  are then
taken into account.   This is  done   by examining  particles  sitting
inside neighboring parent  cells at level $\ell-1$.  Note that in
this case the size of the overlapping cloud is the one of level $\ell$
particles.  In this way,  for a given set  of particle positions, the
resulting density field at level $\ell$ is exactly the same as that of
a regular Cartesian mesh of equivalent spatial resolution.

\subsubsection{Solving the Poisson equation}
\label{poisson}

Several methods   are described  in the literature   to  solve for the
Poisson equation in   the adaptive grids framework  \citep{couchman91,
jessop94, kravtsov97, almgren98, truelove98}. In RAMSES, as in the ART
code, the   Poisson equation is  solved using  a ``one-way interface''
scheme  \citep{jessop94, kravtsov97}:  the  coarse grid solution never
``sees'' the effect  of the fine grids.  The resulting accuracy is the
same as if the coarse grid were alone.  Boundary conditions are passed
only from the coarse grid to the  fine grid by a linear interpolation.
For each AMR level, the solution should  therefore be close to the one
obtained with a Cartesian mesh  of equivalent spatial resolution,  but
\emph{it can not be better in  any way}.  A  better accuracy  would be   
obtained using a  two-way interface scheme,  as the  one described for
example in \citet{truelove98}.   Such a sophisticated improvement
of the Poisson solver is left for future work.

The Poisson equation at the coarse level is solved using standard Fast
Fourier Transform   (FFT)  technique \citep{hockney81},  with  a  Green
function   corresponding  to    Fourier   transform  of    the  $2{\rm
dim}+1$-points finite difference approximation of  the Laplacian.  For
fine levels ($\ell>0$),  the potential  is  found using a   relaxation
method similar  to the  one developed  for the ART  code: the Poisson
equation  is solved using the  $2{\rm dim}+1$-points finite difference
approximation of the Laplacian, with Dirichlet boundary conditions. In
RAMSES, boundary  conditions are defined  in a temporary buffer region
surrounding the level domain,   where the potential is  computed  from
level $\ell-1$ through a linear reconstruction.

Using   these specific   boundary conditions,  the   potential  can be
computed using any   efficient   relaxation method.  In   RAMSES,  the
Gauss-Seidel (GS) method with Red-Black  Ordering and Successive  Over
Relaxation \citep{press92} is used.  In  two dimensions, for unit mesh
spacing, the basic GS writes as
\begin{equation}
\phi^{n+1}_{i,j} = \frac{1}{4}\left(\phi^{n}_{i+1,j}+\phi^{n}_{i-1,j}
+\phi^{n}_{i,j+1}+\phi^{n}_{i,j-1}\right)
- \frac{1}{4}\rho_{i,j}
\end{equation}
This iteration is applied first  to update the potential for ``black''
cells defined  by $i$ odd and $j$  odd or $i$ even   and $j$ even, and
then to update the potential for ``red'' cells defined  by $i$ odd and
$j$   even or $i$   even    and  $j$  odd.   Finally, the   result  is
``over-corrected''   using  the    so-called over-relaxation   parameter
$\omega$
\begin{equation}
\phi^{n+1}_{i,j} = \omega \phi^{n}_{i,j} + (1-\omega) \phi^{n+1}_{i,j}
~~{\rm with}~~1<\omega<2
\end{equation}
The speed of   the algorithm relies on   the correct  choice for  both
$\omega$ and the initial guess $\phi^0_{i,j}$. For a regular $N \times
N$ Cartesian mesh, the optimal over-relaxation parameter is known to be
\citep{press92}
\begin{equation}
\label{optomega}
\omega \simeq \frac{2}{1+\alpha \frac{\pi}{N}}
\end{equation}
where $\alpha=1$ for Dirichlet boundary  conditions and $\alpha=2$ for
periodic boundary conditions. For an irregular AMR grid, the situation
is more   complicated, since the computational  volume  is  covered by
irregular mesh patches.  The over-relaxation  parameter has to be found
empirically. An  interesting way of determining  the optimal value for
$\omega$ is to  estimate the average  size $<L>$ of these patches, and
to use it in formula~(\ref{optomega}) in place of $N$. It was found to
work very well in practice.

The initial guess is obtained  from the coarser level $\ell-1$ through
a linear reconstruction of the  potential.  In this way, the  solution
at  large scale is  correctly  captured at  the  very beginning of the
relaxation process.  Only the shortest wavelengths  need to be further
corrected.

A question  that arises naturally is:  when do we  reach convergence ?
Since our Poisson solver is coupled to a  N-body system, errors due to
the CIC interpolation scheme are dominant in the force calculation. As
soon as the residuals are smaller than the CIC induced errors, further
iterations  are unnecessary.  For   cosmological simulations,  this is
obtained  by specifying that  the  2-norm of the  residual  has to  be
reduced by a factor of at least $10^3$.

Let us consider a $128^3$ coarse grid, completely refined in a $256^3$
underlying fine grid. Solving first the Poisson equation on the coarse
level using FFT, the solution is injected to  the fine grid as a first
guess.  In   this   particular  example,  the  optimal  over-relaxation
parameter  is $\omega    \simeq 1.9$  (using  Eq.~\ref{optomega}  with
$\alpha=2$)   and 60 iterations are     needed  to damped the   errors
sufficiently.

Let us now consider  a more practical example,  in which a typical AMR
grid  is obtained from a cosmological  simulation.  In  this case, the
average AMR patch size was empirically found to be roughly $<L> \simeq
20$  cells.  Equation~(\ref{optomega}) with  $\alpha=1$  gives $\omega
\simeq 1.7$.  20  iterations only are  needed for the iterative solver
to  converge sufficiently.   Note that for   the ART code, the optimal
value was found to be  $\omega=1.25$ \citep{kravtsov97}, using however
a different approach to set up boundary conditions.

\subsubsection{Computing the acceleration}

Using the potential, the acceleration   is computed on the mesh  using
the 5-points finite difference approximation  of the gradient. As  for
the potential,  the   acceleration is cell-centered   and the gradient
stencil  is    symmetrical     in    order    to   avoid    self-forces
\citep{hockney81}. Buffer   regions  defined during  the  previous  
step are  used  here again  to give correct   boundary conditions. The
acceleration is  interpolated back  to  the particles of  the  current
level,  using an inverse CIC  scheme.   Only particles from the linked
list whose cloud  is entirely included  within the  level boundary are
concerned.  For  particles belonging to  level $\ell$, but whose cloud
lies  partially    outside  the level    volume,   the acceleration is
interpolated from  the mesh of level  $\ell-1$.  This is the  same for
the ART code: \emph{``In this way, particles  are driven by the coarse
force until they move sufficiently far into the finer mesh''}
\citep{kravtsov97}.

\subsubsection{Time integration}
\label{timeint}

One requirement in a   coupled N-body and  hydrodynamical code  is the
possibility to deal with variable time steps. The stability conditions
for the time step is indeed given  by the Courant Friedrich Levy (CFL)
condition, which can vary in time.  The standard leapfrog scheme
\citep{hockney81}, though accurate, does  not offer this  possibility.
In RAMSES, a second-order midpoint scheme  has been implemented, which
reduces exactly to the second  order leapfrog scheme for constant time
steps.  Since   the acceleration $-  \nabla \phi^n$  is  known at time
$t^n$  from particle positions  ${\bf  x}_p^n$, positions and velocities
are updated first by a predictor step
\begin{eqnarray}
{\bf v}_p^{n+1/2} = {\bf v}_p^n - \nabla \phi^n \Delta t^n/2
\label{eq:predcorr1} \\
{\bf x}_p^{n+1} = {\bf x}_p^n + {\bf v}_p^{n+1/2}\Delta t^n
\label{eq:predcorr2}
\end{eqnarray}
and then by a corrector step
\begin{eqnarray}
\label{eq:predcorr3}
{\bf v}_p^{n+1}= {\bf v}_p^{n+1/2} - \nabla \phi^{n+1}  \Delta t^n/2
\end{eqnarray}
In this   last  equation,   the  acceleration at   time   $t^{n+1}$ is
needed. In order  to avoid an extra call  to the Poisson solver,  this
last operation is postponed to the next time step. The new velocity is
computed as soon  as the new potential is  obtained. In RAMSES, it  is
possible  to have  either a  single time   step for all  particles, or
individual  time  steps for each  level.  In the  latter case,  when a
particle exits level  $\ell$  with  time step $\Delta  t_{\ell}$,  the
corrector step is applied at  level $\ell-1$, using $\Delta  t_{\ell}$
in place of  $\Delta t_{\ell-1}$. Therefore,  the ``past  history'' of
all particles   has  to be  known in  order   to  apply correctly  the
corrector  step. This   is done  in RAMSES by   introducing one  extra
integer  per particle  indicating  its  current level.  This  particle
``color'' is eventually modified at the end of the corrector step.

Usually, the time     step  evolution  is smooth,   making   our
integration  scheme  second-order in time.  However,   if one uses the
adaptive time   step scheme instead of   the more accurate   (but time
consuming) single time step scheme, the time  step changes abruptly by
a factor of  two for particles crossing a  refinement boundary.   Only
first  order accuracy is  retained  along those particle trajectories.
This  loss of  accuracy has been   analyzed in  realistic cosmological
conditions \citep{kravtsov99, yahagi01} and turns out  to have a small
effect on the particle distribution, when  compared to the single time
step case.

\subsection{Hydrodynamical Solver}
\label{hydrosolver}

In RAMSES, the Euler equations are solved in their conservative form:
\begin{eqnarray}
\label{euler1}
\frac{\partial \rho}{\partial t} + \nabla \cdot \left( \rho {\bf u} 
\right) = 0\\
\label{euler2}
\frac{\partial}{\partial t} \left( \rho {\bf u} \right) + 
\nabla \cdot \left( \rho {\bf u} \otimes {\bf u} \right) + \nabla p 
= - \rho \nabla \phi \\
\label{euler3}
\frac{\partial}{\partial t} \left( \rho e \right) + 
\nabla \cdot \left[\rho {\bf u} \left( e + p/\rho \right) \right] 
= - \rho {\bf u} \cdot \nabla \phi
\end{eqnarray}
where $\rho$ is the mass density, ${\bf u}$ is the fluid velocity, $e$
is the specific total energy, and $p$ is the thermal pressure, with
\begin{equation}
p = (\gamma-1)\rho (e - \frac{1}{2}u^2)
\end{equation}
Note that  the energy equation (Eq.~\ref{euler3}) is conservative
for the  total fluid energy, if  one ignores the  source terms  due to
gravity.  This property is  one of the  main advantages of solving the
Euler equations in conservative form: no energy  sink due to numerical
errors can alter the flow dynamics.  Gravity is included in the system
of equation as a {\em non stiff source term}. In this case, the system
is  not  explicitly conservative  and   the total energy (potential  +
kinetic) is conserved at the percent level (see section \ref{econs}).

Let $U^{n}_{i}$ denote a  numerical approximation to the cell-averaged
value  of $(\rho, \rho {\bf  u}, \rho e)$  at time $t^n$  and for cell
$i$.  The  numerical  discretization   of the  Euler   equations  with
gravitational source terms writes:
\begin{equation}
\frac{U^{n+1}_{i} - U^{n}_{i}}{\Delta t} + 
\frac{F^{n+1/2}_{i+1/2} -F^{n+1/2}_{i-1/2}}{\Delta x} = S^{n+1/2}_{i}
\end{equation}
The  time centered  fluxes $F^{n+1/2}_{i+1/2}$ across  cell interfaces
are  computed  using a  second-order   Godunov method (also  known  as
Pieceweise Linear Method), with or without directional splitting
(according to the   user's choice), while gravitational source  terms
are included using a time centered, fractional step approach:
\begin{equation}
S^{n+1/2}_{i} = \left( 0, 
\frac{\rho^{n}_{i} \nabla \phi^{n}_{i} + 
\rho^{n+1}_{i} \nabla \phi^{n+1}_{i} }{2}, 
\frac{ (\rho u)^{n}_{i} \nabla \phi^{n}_{i} + 
(\rho u)^{n+1}_{i} \nabla \phi^{n+1}_{i} }{2} \right)
\end{equation}   
A general  description of Godunov and  fractional step  methods can be
found in \citet{toro97}.  The  present implementation is based on  the
work of    \citet{collela90} and   \citet{saltzman94}.  For  sake   of
brevity, only its basic features are recalled here.

\subsubsection{Single grid Godunov solver}

In this section, I describe  the  basic hydrodynamical scheme used  in
RAMSES  to solve   equations~(\ref{euler1}-\ref{euler3})   at a  given
level.    It  is assumed that  proper   boundary  conditions have been
provided:  the  hydrodynamical scheme requires 2  ghost  zones in each
side and in each direction, even in the diagonal directions.  Since in
RAMSES the Euler equations  are solved on  octs of $2^{\rm dim}$ cells
each, $3^{\rm dim}-1$ similar neighboring  octs are required to define
proper   boundary conditions.  The basic  stencil   of the PLM  scheme
therefore contains $6^{\rm dim}$ cells.  This is not the case for PPM
\citep{collela84} for  which 4 ghost zones are   required in each side
and in each direction.  Since the AMR structure in  RAMSES is based on
octs ($2^{\rm dim}$ cells), PPM would  be to expensive to implement in
many aspects. One  solution would be to  modify the basic tree element
and increase the number  of cells per  oct from $2^{\rm dim}$ cells to
$4^{\rm  dim}$ cells. The resulting AMR  structure would however loose
part of its flexibility to adapt itself to complex flow geometry.  The
FLASH  code \citep{fryxell00} is an  example of such an implementation,
using the  PPM scheme  in a  similar  recursive tree   structure, with
however $8^{\rm dim}$ cells per basic tree element.

For a given time step,  we need to compute second-order, time-centered
fluxes at  cell interfaces.   This is  done in RAMSES  using a Riemann
solver,  with left and  right  states  obtained  by  a characteristics
tracing step.  A  standard characteristic analysis is done first,
by Taylor expanding the wave  equations to second order and projecting
out the  waves that    cannot reach  the   interface within  the  time
step. These states are then adjusted to  account for the gravitational
field.  If the chosen scheme is not directionnaly splitted, transverse
derivative terms are finally added to account for transverse fluxes
\citep{saltzman94}.  The slopes that enter  into the Taylor  expansion
are computed  using the Min-Mod  limiter to ensure the monotonicity of
the solution.

The Riemann  solver used  to compute  the  Godunov states  is ``almost
exact'',  in    the sense  that a  correct   pressure   at the contact
discontinuity is obtained iteratively (typically, for strong shocks 10
Newton-Raphson  iterations are   required for  single-precision
accuracy  of $10^{-7}$).   The    only approximation relies   in  the
assumption that  the rarefaction wave  has a  linear profile.   In the
final step, fluxes of the conserved variables are computed using these
Godunov states. The outputs of the single grid algorithm are therefore
fluxes across cell interfaces.

Practically, this single  grid module is applied  to a large vector of
stencils of  $6^{\rm dim}$ cells each.   For a large Cartesian grid of
$N^{\rm dim}$ cells, the CPU time overhead associated to this solution
is rather large.  Since  the main time consuming  part is the  Riemann
solver, the estimated CPU time overhead was found  to be roughly 50~\%,
100~\% and 200~\% for ${\rm dim}$=1, 2 and 3 respectively. Since in any
useful AMR calculation, the mesh  structure is not a regular Cartesian
grid,  the  actual  overhead  is much   lower, although   difficult to
estimate  in  practice.  Moreover,  this solution  is   much easier to
implement  than any potentially faster  alternative  one can think of,
and easy to optimize on vector and parallel supercomputers.

\subsubsection{AMR implementation}
\label{timestepping}

This section describes how the solution is advanced in time within the
present AMR methodology.  Note  that this procedure is  recursive with
respect to level $\ell$ (step~3).
\begin{enumerate}
\item Generate new refinements at level $\ell+1$ by
conservative interpolation of level $\ell$ variables.
\item Compute the new time step $\Delta t_{\ell}$ using the CFL
Courant condition and the constraint $\Delta t_{\ell} \le
\Delta t_{\ell-1}$.
\item Advance the solution in time for level $\ell+1$, once in the single 
time step case, or twice for the adaptive time step case.
\item Modify the time step $\Delta t_{\ell}$ according to the 
synchronisation constraint $\Delta t_{\ell} = \Delta t_{\ell+1}$ 
for the single time step case or $\Delta t_{\ell} = 
\Delta t_{\ell+1}^1 + \Delta t_{\ell+1} ^2$ for the adaptive time step case.
\item Compute boundary conditions in a temporary buffer
by conservative interpolation of level $\ell-1$ variables.
\item Compute fluxes using the single grid Godunov solver.
\item Replace the fluxes at coarse-fine interface by averaging the fluxes
computed at level $\ell+1$.
\item For leaf cells, update variables using these fluxes.
\item For split cells, update variables by averaging down the updated variables
of level $\ell+1$.
\item Build the new refinement map.
\end{enumerate}
In RAMSES,  boundary  conditions  are supplied  to  fine  levels by  a
conservative linear reconstruction of coarse cell values (step~5). The
actual  interpolation  scheme is a   3D generalization of  the Min-Mod
limiter  \citep{dezeeuw93}. The coarse  solution  is assumed to remain
constant in time during  the advance of  the fine solution.  For  fine
cells at  coarse-fine boundaries and for the adaptive time step case
only,  the accuracy reduces thus from  second to  first order in time,
but the global solution remains second order \citep{khokhlov98}.

\subsection{Time Step Control}

The  time  step is  determined   for each  level independently,  using
standard  stability constraints  for  both N-body  and  hydrodynamical
solvers.

The first   constraint comes from  the gravitational  evolution of the
coupled  N-body   and hydrodynamical   system, imposing   that $\Delta
t^{\ell}$ should be  smaller  than a  fraction $C_1<1$  of the minimum
free-fall time
\begin{equation}
\Delta t_1 ^{\ell} = C_1 \times \min _{\ell}(t_{{\rm ff}})
\end{equation}
An additional constraint comes from  particle dynamics within the  AMR
grid, imposing that particles move  by only a  fraction $C_2<1$ of the
local cell size.
\begin{equation}
\Delta t_2^{\ell} = C_2 \times \Delta x ^{\ell}/\max _{\ell}(v_p)
\end{equation}
A third constraint is imposed on the time step  by specifying that the
expansion factor $a_{\rm exp}$  should not vary  more than $C_3 \simeq
10~\%$ over  one time  step.  This constraint  is active  only at early
times, during the linear regime of gravitational clustering.
\begin{equation}
\Delta t_3^{\ell} = C_3 \times a_{\rm exp} / \dot{a}_{\rm exp}
\end{equation}
The last constraint is imposed by the Courant Friedrich Levy stability
condition, which states that the time step should be smaller than
\begin{equation}
\Delta t _4^{\ell} = {\rm cfl}\times 
\Delta x^{\ell} / \max_{\ell} (|u_x| + c, |u_y| + c, |u_z| + c)
\end{equation}
where ${\rm cfl} < 1$ is the Courant factor. In the coupled N-body and
hydrodynamics  case,  the actual time  step  is equal to $\min (\Delta
t_1^{\ell}, \Delta t _2^{\ell}, \Delta t _3^{\ell}, \Delta t _4^{\ell})$.

\subsection{Refinement Strategy}
\label{sectrefinecriteria}

The refinement strategy is the key issue for any AMR calculation.
Bearing in mind that the overhead associated  to the AMR scheme can be
as  large as  a  factor of  2  to  3 (see  section  \ref{hydrosolver})
compared to  the  corresponding  uniform grid  algorithm,  the maximum
fraction of the grid that can be refined lies in between 30~\% to 50~\%.
One should  therefore design a refinement  strategy that allows for an
accurate  treatment of the underlying  physical problem, but minimizes
also the fraction of the volume to be refined. 

For   the  N-body solver, the   refinement  strategy is  based  on the
so-called ``quasi-Lagrangian''  approach.   As in  \citet{kravtsov97},
the idea is  to obtain  a constant number  of particles  per cell.  In
this way, two-body relaxation effects can be minimized, as well as the
Poisson noise due to particle discreteness effects.  The latter effect
can be  damaging   when  coupling the    N-body code   to  the
hydrodynamics  solver.     The ``quasi-Lagrangian''    approach     is
implemented by  refining  cells at  level  $\ell$ if  the  dark matter
density exceeds a level dependent density threshold, defined as
\begin{equation}
\rho_{\ell} = M_c \times (\Delta x^{\ell})^{-\rm dim}
\end{equation}
where $M_c$  is the maximum  mass (or  number of  particles) per cell.
For pure N-body  simulations,   $M_c$ is  usually chosen  around  5-10
particles \citep{kravtsov97}, which gives  a few particles per cell on
average.  For gas dynamics simulations, $M_c$  should be chosen around
40-80 particles, in order to lower  enough the Poisson noise.  In this
case,  we  obtain  indeed   more   than  10  particles  per  cell   on
average. Note also that since  for gas dynamics simulations, the total
memory is dominated by the storage associated  to the fluid variables,
the number of  particles par cell can be  chosen much higher  than for
pure dark matter simulations.

As   for the N-body  solver,  a  ``quasi-Lagrangian'' approach can  be
implemented  for  the  gas  component, using   level dependent density
thresholds defined by
\begin{equation}
\rho_{\ell} = M_b \times (\Delta x^{\ell})^{-\rm dim}
\end{equation}
In order  to follow the same Lagrangian  evolution than the dark matter
component, the typical baryonic mass per cell $M_b$ can be derived as
\begin{equation}
M_b = M_c \frac{\Omega_b}{\Omega_m-\Omega_b}
\end{equation}
For pure gas dynamics applications,  other refinement criteria can  be
used \citep[see][  for more  examples]{khokhlov98}.   In  RAMSES, only
refinement criteria based on gradients of the flow variables have been
implemented: for each cell $i$ and for  any relevant flow variable $q$
(pressure, density, Mach number...), its gradient is computed using the
$2\times \rm dim$  neighboring cells.  If   this gradient, times  the
local mesh spacing, exceeds a fraction of the central cell variable
\begin{equation}
\nabla q _i \ge (\nabla q)_{\max}^{\ell}
 = C_{q}\frac{q_i}{\Delta x^{\ell}}
\end{equation}
then cell  $i$ is refined.  The parameter $C_{q}$  is a free parameter
that need to be specified by the  user. A similar criterion based
on  second  derivatives   of    the  flow variables   has   also  been
implemented.

The last refinement criterion implemented in RAMSES is purely spatial:
for each level, refinements are not  allowed outside a sphere centered
on the box center.  This last criterion allows the  user to refine the
computational mesh only in the  center of the  box, in order to follow
properly the formation of a single structure, without spending to much
resources in  refining  also the surrounding   large  scale field. The
radius of this spherical  region,  noted $R_{\ell}$, can be  specified
for each level independently.

\subsection{Cosmological Settings}
\label{supercomoving}

RAMSES  can be  used for  standard fluid dynamics  or N-body problems,
with periodic, reflecting, inflow  or outflow boundary conditions.  For
the present paper, RAMSES is however used in the cosmological context.
The N-body solver and  the  hydrodynamics solver are both  implemented
using  ``conformal time''  as   the  time  variable.  This  allows   a
straightforward  implementation of   comoving coordinates, with  minor
changes to the  original  equations.  The  details of  these so-called
``super-comoving coordinates''   can be found in  \citet{martel98} and
references therein.  The  idea is to perform  the  following change of
variables
\begin{equation}
d\tilde t = H_0 \frac{dt}{a^2}
~~~{\rm and}~~~
\tilde x = \frac{1}{a} \frac{x}{L}
\end{equation}
\begin{equation}
\tilde \rho = a^3 \frac{\rho}{\Omega_m \rho_c}
~~~{\rm and}~~~
\tilde P = a^5 \frac{P}{\Omega_m \rho_c H_0^2 L^2 }
\end{equation}
\begin{equation}
\tilde {\bf u} = a \frac{{\bf u}}{H_0 L}
\end{equation}
where $H_0$  is the Hubble constant, $\Omega_m$  is the matter density
parameter, $L$ is  the box size  and $\rho_c$ is the critical density.
In  the specific  case  $\gamma =  5/3$,  equations (\ref{nbody})  and
(\ref{euler1}-\ref{euler3}) remain unchanged, at  the exception of the
Poisson equation which now reads
\begin{equation}
\triangle_x \tilde \phi = \frac{3}{2} a \Omega_m (\tilde \rho - 1)
\end{equation}
If $\gamma \neq 5/3$, a single additional source term must be included
in   the   right-hand   side of  the     energy conservation  equation
(Eq.~\ref{euler3}).     These  ``super-comoving coordinates''  simplify
greatly the introduction of comoving variables in the equations.

\section{Tests of the code}
\label{sec:tests}

In this section, I present tests of increasing complexity for both the
N-body solver  and the hydrodynamical  solver.   These tests  are also
useful to  choose the correct   parameters for realistic cosmological
applications described in the last section.

\subsection{Acceleration around a Point Mass}

\begin{figure*}
\centering
\caption{
Acceleration of massless test particles dropped randomly around single
massive  particles, whose positions  are also  chosen randomly  in the
box. The coarse grid has $32^3$ cells. The number of refinement levels
is  progressively  increased  from  0  to 6,  with  increased  spatial
resolution around the massive  particles.  The radial AMR acceleration
divided by the  true acceleration is shown as  light grey dots (versus
radius  in  units  of coarse  cell  size).   The  same ratio  for  the
tangential AMR  acceleration is  shown as dark  grey dots.   The force
corresponding  to  an homogeneous  sphere  with  radius  equal to  the
smallest cell length is also plotted for comparison (solid line).  }
\label{fig1}
\end{figure*}

Particles of  unit mass are placed randomly  in the computational
box, whose coarse grid  is defined by $n_{\mathrm{x}} = n_{\mathrm{y}}
=  n_{\mathrm{z}}=32$.  Test  particles are  then dropped  randomly in
order to  sample the acceleration  around each massive  particle.  The
AMR grid  is built  around each central  particle.  For  that purpose,
refinement  density thresholds  were set  to $\rho_{\ell}=0$  for each
level.   An increasing  number  of refinement  levels  was used,  from
$\ell_{\mathrm{max}}=0$  to $\ell_{\mathrm{max}}=6$,  the  latter case
corresponding to a formal  resolution of $2048^3$.  Mesh smoothing was
performed with $n_{\mathrm{expand}}=1$.

Figure~\ref{fig1}   shows   the   resulting  radial   and   tangential
accelerations,  divided by  the  true $1/r^2$  force.  The  tangential
acceleration gives here an indication of the level of force anisotropy
and accuracy.  Note  that the acceleration due to  the ghost images of
the  massive particle (periodic  boundary conditions)  was substracted
from  the computed  acceleration (using  the Ewald  summation method).
For comparison, the acceleration of an homogeneous sphere (with radius
equal  to the  cell  size of  the  maximum refinement  level) is  also
plotted in  Figure~\ref{fig1} as a  solid line : the  AMR acceleration
appears  to provide a  slightly lower  spatial resolution  (rouhly 1.5
cell size).  At lower radius, the force smoothly goes to zero, exactly
as for a PM code of  equivalent dynamical range.  At lower radius, the
force  anisotropy is  also the  same as  for a  PM code  of equivalent
dynamical  range.  On  the other  hand,  at higher  radius, the  force
anisotropy  remains close  to  1~\%.   Contrary to  a  single grid  PM
solver,  the force  error does  not decrease  monotonically  as radius
increases.   Here, the error  level remains  roughly constant  (at the
percent level), since the  spatial resolution also decreases as radius
increases. In fact, the AMR force on a given particle corresponds to a
single  grid  PM  force whose  cell  size  is  equal  to that  of  the
particle's current  level.  As  one goes from  one level to  the next,
discontinuities in the force remain also at the percent level.

\subsection{Acceleration of Particles in a $\Lambda$CDM Simulation}

\begin{figure*}
\centering
\caption{
Particle   positions   obtained   in a   $\Lambda$CDM   simulation are
considered  in   this test.  Each   panel shows  the force   error for
particles sitting  at different  levels of   refinement. The error  is
defined as the difference between the AMR force and the force computed
by  a  PM code of  equivalent  spatial resolution.  In each panel, the
average and the variance of the error are also shown. }
\label{fig3}
\end{figure*}

In  order to  assess  the quality   of the gravitational  acceleration
computed by RAMSES in a cosmological situation, we  consider now a set
of $64^3$ particles obtained in a $\Lambda$CDM simulation, at redshift
$z=0$. In this  way,  we are able  to  quantify the force errors  in a
typical hierarchical  clustering configuration, with the corresponding
mesh  refinements  structure.     The coarse  grid    was defined   by
$n_{\mathrm{x}}=n_{\mathrm{y}}=n_{\mathrm{z}}=32$  and  each  particle
was assigned  a mass $m_p  = 1/8$. The  adaptive mesh  was built using
refinement density thresholds $\rho_{\ell}=5 \times 8^{\ell}$ for each
level $\ell$. Each cell is therefore refined  if it contains more than
40 particles, with a roughly   constant number of particles per   cell
after  each refinement (between  5 and 40). Mesh structures associated
to this  particle distribution  are shown  in the last  section of the
paper (Fig.~\ref{figlargescale}).

For each level of refinement, the AMR force is then compared to the PM
force of  equivalent  spatial  resolution (see  Fig.~\ref{fig3}).  For
particles sitting  at  the coarse level  ($\ell=0$),  the force is  by
construction exactly equal to the PM force  with $32^3$ cells (results
not shown in the figure).  For levels $\ell=1$, $\ell=2$, $\ell=3$ and
$\ell=4$, the AMR force is compared  to the PM force with respectively
$64^3$, $128^3$, $256^3$ and $512^3$ cells. In Figure~\ref{fig3}, each
panel shows the difference between the AMR force and  the PM force for
each level. The number of particles sitting at each level is indicated
in the upper-left  part of each  panel. The mean  force error and  the
standard deviation  is   indicated  in the  lower-left  part   of each
panel. Although the error  distribution is strongly non  Gaussian, its
typical magnitude remains at   the percent level  in all  cases.  Note
that errors  are larger for forces  of intermediate and  small values,
indicating that  those  particles might be  sensitive  to the boundary
conditions (tidal field) imposed on the level  boundaries (this is the
main  source of inaccuracy in the  N-body scheme).  On the other hand,
for  particles with strong   acceleration,  the  AMR force is   almost
indistinguishable from the PM force of equivalent resolution.

\subsection{Shock Tube}

\begin{figure*}
\caption{
Shock Tube Test: Density, velocity, pressure and refinement level as a
function of position at time $t=0.245$. Numerical results are shown as
squares,  and compared to the analytical  solutions (solid lines). See
text for details.}
\label{fig4}
\end{figure*}

The initial   conditions are  defined     by a left  state  given   by
$\rho_L=1$, $u_L=0$    and $P_L=1$ and  by a    right  state given  by
$\rho_R=0.125$, $u_R=0$ and  $P_R=0.1$ for a  $\gamma=1.4$ fluid. This
test  (also called Sod's test)  is interesting because it captures all
essential features of one  dimensional hydrodynamical flows, namely  a
shock wave, a contact discontinuity  and a rarefaction wave. While the
latter   wave remains    continuous,  the   2    former features   are
discontinuous.  Modern shocks  capturing methods like  the one used in
RAMSES spread shock fronts over 2 to 3 zones.  Contact discontinuities
are usually more difficult  to capture (say 6 to   10 cells), and  the
spreading  usually  increases  with the  number of    time step.  This
numerical smoothing is responsible  for the dissipation of the scheme.
AMR technique allows one to increase the spatial resolution around the
discontinuities and therefore  to minimize the numerical  dissipation.
In the present   application, the refinement   criteria are based   on
pressure,     density    and       Mach   number     gradients    (see
Sect.~\ref{sectrefinecriteria}),              with           parameter
$C_{\rho}=C_P=C_M=0.01$.  The maximum number of refinements was set to
$\ell_{\rm max}=6$, for a   coarse  level mesh  size $n_x=64$.    Mesh
smoothing   (see \S~\ref{sectrefinemap}) is   performed  using $n_{\rm
expand}=1$.  Note that the refined  mesh is built before the beginning
of the  simulation. The time step is   controlled by a  Courant number
$cfl=0.8$.  Results are shown  at time $t=0.245$  and compared  to the
analytical solution in Figure~\ref{fig4}.    The shock front  and  the
contact  surface are refined up  to the maximum  refinement level: the
formal   resolution is therefore 4096.      The total number of  cells
(counting both split and  leaf  cells) is only  560,  or 14~\% of  the
corresponding uniform mesh size.  69  time steps were necessary at the
coarse level, while 4416 time steps were  necessary at level $\ell=6$.
It is worth  mentioning that pressure  and velocity remain  remarkably
uniform  across the contact discontinuity, and  no side effects due to
the presence of discrete refinement ratio are noticeable.

\subsection{Planar Sedov Blast Wave}

\begin{figure*}
\caption{
Planar   Sedov Blast   Wave  Test:  Density,  velocity,  pressure  and
refinement level as  a function of  position for times $t=$0.05,  0.1,
0.15, 0.2, 0.25 and 0.3.  Numerical results are  shown as squares, and
compared to the  analytical  solutions  (solid  lines). See  text  for
details.}
\label{fig5}
\end{figure*}

The last test,   though  interesting and  complete,  is  not  a  very
stringent one, since it involves a rather weak shock. In order to test
the ability  of RAMSES to handle  strong  shocks (a  common feature in
cosmology),     let  us  consider  the   planar    Sedov  problem: the
computational  domain   is   filled with   a  $\gamma=1.4$  fluid with
$\rho_0=1$,   $u_0=0$ and $P_0=10^{-5}$.   A   total (internal) energy
$E_0=1/2$ is  deposited in the first  cell only  at $x=0^+$. Note that
here again  the  refined mesh is   built before the beginning   of the
simulation. Reflexive boundary  conditions are considered. The grid is
defined by  $n_x=32$ with 6 levels  of refinement. The only refinement
criterion    used  here  is    based  on   pressure   gradients,  with
$C_P=0.1$.  Mesh  smoothing is guaranteed  by  $n_{\rm expand}=1$. The
Courant number is set to $cfl=0.8$.  Very rapidly, a self-similar flow
builds  up,   following    the  analytical   solution   described   in
\citet{sedov93}.   Simulation results  are  shown for different output
times  and compared to the analytical  solutions. Note  that the shock
front propagates exactly at the correct  speed. The numerical solution
closely matches the  analytical one,  without any  visible  post-shock
oscillations.  239 time steps only were necessary at the coarse level,
but 15296  time steps were necessary  at  the finest refinement level.
The total number of cells (including split cells) in the adaptive mesh
structure is only 130, to be compared  with 2048 cells for the uniform
grid of equivalent spatial resolution (4.3~\%).  Due to the refinement
criterion used    here,  the adaptive   mesh  mainly  concentrates the
computational effort around the shock  front.  In one dimension, as it
is the case here, discontinuities like shocks are quite inexpensive to
deal with:  if one adds one level  of refinement, the total  number of
cells  increases by  a   constant (and  small) amount.   For  two- and
three-dimensional calculations,  the situation is much more demanding:
since  shocks  and  contacts    discontinuities are  surface    waves,
increasing  the resolution by a  factor of 2 corresponds to increasing
the total number of cells by a factor of 2 for ${\rm dim}=2$ and 4 for
${\rm dim}=3$. Therefore, we have  to face the possibility of stopping
at some level the refinement hierarchy and investigate what happens to
the numerical solution in this case.

\subsection{Strong Shock Passing through a Coarse--Fine Interface}
\label{shockcoarsetofine}

\begin{figure}
\caption{
Strong Shock Passing through a Coarse--Fine Interface with Mach number
$M=5000$  and  $\gamma=5/3$, computed with   $cfl=0.5$. The upper plot
shows the reference case  with   a 256  cells coarse grid,   uniformly
refined up to level  $\ell=1$ (without any coarse--fine boundary). The
middle    figure shows  the   case  where the  shock   goes  through a
fine--to--coarse boundary (located  around $x \simeq 234$),  while the
bottom   figure  shows    the   case where   the   shock    goes to  a
coarse--to--fine boundary  (located  around the  same  place). In  the
latter case,  perturbations of the order  of 5~\%  are generated in the
post--shock flow.}
\label{fig4b}
\end{figure}

It is well known that in any  AMR calculations, discontinuities in the
flow (like the one discussed in the previous sections) must be refined
up to  the maximum level,    in  order  to  obtain accurate    results
\citep{berger89,khokhlov98}. Unfortunately, it  is not always possible
to   satisfy this rule because of   memory limitations, even on modern
computers.  One  has      therefore to  consider  cases   for    which
discontinuities leave   or    enter  regions  of     different spatial
resolution.  The    situation  is  especially   sensitive  for contact
surfaces, since as soon  as the code spreads  them over, say, 6 cells,
no  matter how much  one  refines them  afterwards, they will preserve
their original thickness. Shock waves, however, have a self-steepening
mechanism that   allows  them to adapt  to   the local  resolution and
restore their sharp profile over 2 to 3 cells only.   The price to pay
for this     interesting property is   the  appearance   of post-shock
oscillations after the front has entered  a high-resolution region. To
illustrate this,  \citet{khokhlov98} proposed a  simple  test based on
the  propagation of   a  strong  shock   wave across  a   coarse--fine
interface.   Khokhlov's  test is reproduced   here using the following
parameters:   the  shock  Mach    number is  set  to   $M=5000$   with
$\gamma=5/3$. The  base grid  resolution is set  to $n_x=256$  and the
Courant number is set to $cfl=0.5$.

The following three cases have been considered:
\begin{enumerate}
\item the whole computational domain is refined up to $\ell=1$.
\item the computational domain is refined up to $\ell=1$ only left to
$x=234$.
\item the computational domain is refined up to $\ell=2$ right to
$x=234$ and up to $\ell=1$ otherwise.
\end{enumerate}
The resulting density profiles are shown in Figure~\ref{fig4b}.  While
in  the 2   former   cases, the   density profiles   show  no  visible
oscillations,  the latter case does  show  small  oscillations of  the
order of  5~\%. This is  a direct consequence  of the abrupt change of
spatial  resolution between the  2 levels of refinement \citep[see the
discussion  in][]{berger89}. To summarize,   if shock waves move  from
high- to  low-resolution regions, spurious  effects associated  to the
mesh structure are undetectable. This is not  the case in the opposite
situation,   which   causes    spurious (though    small)   post-shock
oscillations.  Note   however  that for weak    shocks  the  effect is
undetectable \citep{berger89}.  In cosmology,  it is  worth mentioning
that, since the basic features are accretion  shocks, we are always in
a  favorable  situation:  strong   shocks originate    in high-density
(high-resolution) regions, and propagates  outwards, in a  low-density
(low-resolution) background. To  my opinion, this fundamental property
allows  us  to  use safely  adaptive   mesh  technique  in cosmological
simulations.

\subsection{Spherical Sedov Blast Wave}

\begin{figure*}
\caption{
Spherical Sedov Blast Wave  Test: Rescaled density, velocity, pressure
and volume-averaged refinement level as a function of radius for times
$t=10^{-4}$, $5.7\times  10^{-4}$ and $3.2  \times 10^{-3}$. Numerical
results (solid lines with error bars) are compared to the
analytical solutions (solid lines). See text for details.}
\label{fig6}
\end{figure*}

We now consider a very difficult test for Cartesian grids like the one
used in  RAMSES: the spherical Sedov test.  In contrary to the planar,
1D  case, the spherical blast  wave is now fully three-dimensional and
pretty far from the natural geometry  of the code. Moreover, as stated
before, shock waves  in 3D are  essentially two-dimensional: the total
number of cells necessary to cover the shock front scales with spatial
resolution as
\begin{equation}
N_{\rm cells} \propto \left( \frac{R_s}{\Delta x} \right) ^{{\rm
dim}-1}
\end{equation}
where $R_s$ is the curvature  radius of the  shock. For ${\rm dim}=3$,
one  clearly    sees that   the   number  of  required  cells  quickly
explodes. On the other hand, for the Sedov blast wave test, it is more
interesting to   keep the {\em relative}  thickness   of the shock low
enough to   capture the true solution.   As  we have seen  in the last
section,   if  one degrades  the  resolution as   the shock propagates
outwards, no spurious effects are expected.  In RAMSES, we can enforce
a position--dependent spatial resolution by  forbidding a given  level
of refinement to be activated if the radius of the cell is larger than
a   given   threshold  (see   \S~\ref{sectrefinecriteria}).    The run
parameters  are therefore the followings: the  coarse grid size is set
to $n_x=n_y=n_z=32$ and the maximum level  of refinement is chosen to
be $\ell_{\rm max}=6$. The maximum refinement radius for each level is
given by
\begin{equation}
\label{maxrad}
R_{\ell} = 2^{5-\ell} ~~~ {\rm for } ~~~ 0 \le \ell \le 5
\end{equation}
We use refinement criteria  based on pressure gradients with $C_P=0.5$
and mesh smoothing  parameter   $n_{\rm expand}  =  1$. The  fluid  is
supposed initially at rest  with $\rho_0=1$ and $P_0=10^{-5}$. A total
(internal) energy  $E_0=1$ is   deposited   in  the 8  central   cells
only. The refined mesh is here again built before the beginning of the
time integration. We assume $cfl=0.8$ and  $\gamma=1.4$. We use here a
single time  step for all levels,  since in this particular case, this
is the fastest solution.

Results are shown in  Figure~\ref{fig6} and compared to the analytical
solutions  of  \citet{sedov93} for  3  different  output times.   Each
quantity represents a volume-average  value over spherical bins, whose
thicknesses correspond to the local resolution. Each quantity was also
rescaled  according  to  the  (time-dependent)  analytical  post-shock
values (labeled with  an ``$s$'') for sake of  visibility.  Error
bars  are  computed using  the  standard  deviation  of the  numerical
solution from  the mean value  in each spherical bin.   The agreement
with the analytical solution  is remarkable, considering that the mesh
has a Cartesian geometry.  The main departure is found in the velocity
profile at a radius around  60~\% of the shock radius. Similar results
were  obtained by  \citet{fryxell00}.   An easy  way  of solving  this
problem  would be  to lower  the pressure  gradients  threshold $C_P$,
which would  directly increase the  resolution in this region,  at the
expense of increasing the total  number of cells.  Oscillations due to
spurious  mesh reflections  are not  visible in  the  radial profiles.
Direct inspection of the 3D data shows that the only systematic effect
is the departure  from spherical symmetry due to  the Cartesian nature
of  the mesh.   In Figure~\ref{fig6},  the  volume-averaged refinement
level is  shown as a function  of radius for different  times.  Due to
the maximum refinement radii  we used (Eq.~\ref{maxrad}), the adaptive
mesh evolution  is also self-similar,  though in a  piecewise constant
manner.  The  total number  of cells (and  therefore the  memory used)
remains roughly  constant over  the calculation (around  $10^6$ cells,
including split  cells).  The interest of using  a tree-based approach
for building the adaptive mesh appears clearly in this test, since the
mesh structure follows  as closely as possible the  spherical shape of
the shock front.

\subsection{Zel'dovich Pancake}

\begin{figure*}
\caption{
Zel'dovich Pancake  Test:  Density, velocity, pressure  and refinement
level as a function  of position from  the pancake mid-plane at  $z=0$.
The solid  line shows AMR results  if refinements are  activated using
both density thresholds and  pressure gradients. This explains why the
2  accretion   shocks are refined.  The  squares  show  AMR results if
refinements are activated using only  density thresholds. See text for
details.}
\label{fig8}
\end{figure*}

In  this  section,  typical   conditions encountered  in  cosmological
simulations are addressed using the Zel'dovich  pancake test. This test
is   widely  used   to benchmark    cosmological  hydrodynamics  codes
\citep{cen92,ryu93,bryan97,teyssier98}, since  it encompasses  all
the relevant physics (gravity, hydrodynamics and expansion). It can be
thought as  a single mode  analysis of the  collapse of random density
perturbations,   a  first  step   towards  the   study of   the  fully
three-dimensional case. The initial conditions are defined for a given
starting  redshift   $z_i$    in   an  Einstein-de   Sitter   universe
($\Omega_m=1$,   $\Omega_b=0.1$),   using    a     sinusoidal  density
perturbation  of  {\em unit  wavelength}, i.e. of the form
\begin{equation}
\frac{\delta \rho}{\rho} = \frac{1+z_c}{1+z_i}
\cos \left( 2\pi x \right)
\end{equation}
where $x$ is the comoving distance to  the pancake mid-plane (from now
on,   we  always   use  super-comoving   coordinates,  as  defined  in
\S~\ref{supercomoving}). The  initial velocity  field is set according
to the linear theory of gravitational instability
\begin{equation}
u = - \frac{1}{2\pi} \frac{1+z_c}{(1+z_i)^{3/2}}
\sin \left( 2\pi x \right)
\end{equation}
The collapse  redshift is  chosen to be  $1+z_c=10$  and the  initial
redshift is $1+z_i=100$. The initial baryons temperature  was set to a
very low arbitrary  value,   consistent with a  negligible  background
temperature.

The coarse grid is defined by $n_x=32$. We use $N_p=256$ particles for
the dark matter component. The maximum  level of refinement was set to
$\ell_{\rm max}=6$, corresponding to a  formal resolution of 2048. Two
different   cases are investigated:  in  the first  run, both pressure
gradients and gas density thresholds  (quasi-Lagrangian mesh) are used
to build the adaptive mesh, with
\begin{equation}
\rho_{\ell} = 2 \Omega_b \rho_c \frac{\Delta x_{0}}{\Delta x_{\ell}}
~~~{\rm and }~~~C_P=0.5
\end{equation}
while in  the second case,  only gas  density  thresholds are used  to
trigger new refinements.

Results are shown in Figure~\ref{fig8} for  both cases.  For the first
case,  the two accretion shock  fronts are refined   up to the maximum
refinement level, and are therefore very sharp.   For the second case,
however, the shock fronts are not refined at all.  The shock waves are
traveling outwards, from   the high-resolution region  in the  pancake
center, to the  low-resolution background. In  light of what have been
discussed in the previous sections, this  explains why no oscillations
(due to  potential   spurious  reflections at level    boundaries) are
visible.  It is worth mentioning that both sets of profiles are almost
indistinguishable in the center of the pancake. This last test is very
encouraging,  since it  allows us  to  avoid refining  shock fronts in
cosmological  simulations.  The   opposite  situation would  have been
dramatic, because of  the large filling  factor of cosmic shock  waves
(especially in  3D),  which  would  result  in   a very large   memory
overhead, and because it  would  trigger  collisionality in  the  dark
matter particles distribution.

\subsection{Spherical Secondary Infall}
\label{sectsecinfall}

\begin{figure*}
\caption{
Secondary  Infall  Test:  Rescaled  density,  velocity,  pressure  and
volume-averaged refinement level as a  function of radius (in units of
coarse  cells length)  for expansion  factors $a=64a_i$,  $512a_i$ and
$4096a_i$.  Numerical  results  (solid  lines  with  error  bars)  are
compared  to the  analytical solutions  of Bertschinger  (1985) (solid
lines).  See text for details. }
\label{fig9}
\end{figure*}

The last test, while  interesting, is not  very stringent, since it is
very   close  to the   natural, Cartesian  geometry   of  the code. An
analytical  solution  describing   the  fully  non-linear  collapse of
spherical  density perturbations was  found by \citet{bertschinger85},
for  both pure  dark matter   and pure   baryons  fluids. The  initial
conditions  defining the   system  are the  followings:  a  completely
homogeneous Einstein-de Sitter universe (with $\Omega_m=1$) contains a
single   mass   perturbation $\delta    M_0$  at some    initial epoch
$t_0$. Surrounding this initial seed, shells of matter with increasing
radius starts expanding within  the Hubble flow, but finally decouples
from the expansion at some ``turn around'' time, and the corresponding
turn around {\em proper} radius, given by
\begin{equation}
r_{ta}(t) = \left( \frac{4}{3\pi} \frac{t}{t_i} \right)^{8/9}
\left( \frac{3}{4\pi}\frac{\delta M_0}{\rho_c} \right)^{1/3}
\end{equation}
Since in the problem, no other time- or length-scale are involved,
the overall evolution is self-similar.

In  \citet{kravtsov97},  the  secondary  infall test  was successfully
passed by the ART code for the pure dark matter case. Results obtained
by RAMSES are very close to the  ones obtained by  the ART code, which
is  reassuring,  since   both codes   have almost   the  same  N--body
solver. They are not presented here. Rather, we investigate the purely
baryonic  self--similar infall, so   as to validate  the hydrodynamics
solver coupled to gravity and cosmological expansion.

The periodic box is initially filled  with a critical density cold gas
with $\gamma=5/3$.  A single dark  matter particle with mass $m_p=1/8$
is   placed as  initial   seed  in the  center  of  the  computational
domain. The coarse grid is defined by $n_x=n_y=n_z=32$ and the maximum
level of  refinement was set   to $\ell_{\rm max}=6$,  providing us  a
formal  spatial resolution of $2048^3$.   Before the beginning of time
integration, the mesh  is  refined  around the  central seed   using a
maximum refinement radius for each level given by
\begin{equation}
R_{\ell} = 2^{3-\ell} ~~~ {\rm for } ~~~ 0 \le \ell \le 5
\end{equation}
in units of coarse cell size. The resulting mesh structure can be seen
in Figure~\ref{fig9}  in a volume-averaged radial representation, with
roughly   $10^5$ cells  in the AMR    tree, including split cells.  No
pressure  gradients criterion is used,   so that shock fronts are  not
refined. We use   a Courant factor   $cfl=0.5$. Starting at  expansion
factor   $a_i=10^{-5}$, three      output   times   were      analyzed
($a\simeq64a_i$, $512a_i$ and  $4096a_i$). The final epoch was reached
in 86 coarse time steps only, but 5504 time steps at the maximum level
of refinement.

The resulting  {\em rescaled} density, pressure  and velocity profiles
are  plotted  in  Figure~\ref{fig9}  and compared  to  the  analytical
solution of \citet{bertschinger85}. Error bars are computed using
the standard deviation  of the numerical solution with  respect to the
average radial value. The scaling relations for velocity, density and
pressure are obtained using their ``turn around'' values
\begin{equation}
\rho_{ta}(t) = (6\pi Gt^2)^{-1}
\end{equation}
\begin{equation}
u_{ta}(t)=\frac{ r_{ta}(t)}{t}
\end{equation}
\begin{equation}
P_{ta}(t)=\rho_{ta}(t) \frac{ r_{ta}(t)^2}{t^2}
\end{equation}
We  find  a  very  good  agreement between  numerical  and  analytical
profiles,  down to a  radius of  2 fine  cells, the  actual resolution
limit  of the  code. As  the  shock propagates  outwards, no  spurious
reflection appears, as expected.

\section{Application to Cosmology: Structure Formation in a 
$\Lambda$CDM universe.}
\label{sec:simucosmo}

In  this  section,   results obtained  by   RAMSES for  a   N-body and
hydrodynamical simulation  of structure  formation  in a  $\Lambda$CDM
universe are reported. The box  size  was set to $L=100~{\rm  h}^{-1}$
Mpc, as a good compromise between cosmic variance and resolution.  The
influence of the  chosen box size  on the results are not investigated
in this  paper.  On the other hand,  the convergence properties of the
solution are examined by varying the mass and spatial resolution using
6 different runs, whose parameters are listed in Table~\ref{tableruns}
below.  Numerical results  are   also compared to  analytical  results
obtained in  the  framework of  the  halo model.    This simple theory
predicts   various  quantities for     both    gas and   dark   matter
distributions,  and  has already   proven  to successfully   reproduce
results    obtained   in   various     numerical   simulations    (see
Sect.~\ref{sec:halomodel}).      A careful    comparison  between  the
analytical and  the  numerical approach will   serve us as a guide  to
investigate our understanding of structure formation in the universe.

\subsection{Initial conditions}

An    initial Gaussian random   field was  generated   for the highest
resolution  run on a $256^3$  particle grid and  (periodic) box length
$L=100 {\rm h}^{-1}$ Mpc. The  transfer function of \citet{ma98} for a
flat $\Lambda$CDM universe was used  and  normalized to the COBE  data
\citep{white95}, with the following cosmological parameters
\begin{equation}
\Omega_m=0.3 ~~~ \Omega_\Lambda=0.7 ~~~ \Omega_b =0.039
\end{equation}
\begin{equation}
h=0.7 ~~~ \sigma_8 = 0.92
\end{equation}
The high resolution grid was then degraded twice  (down to $128^3$ and
$64^3$)  to provide consistent     initial  conditions for our     low
resolution runs.  In   this way, a  direct  comparison  between the  3
simulations   is  made possible.    The corresponding  mass resolution
(corresponding to individual particle  masses  for a pure  dark matter
universe)   is  $M_0=5\times  10^9$  $M_{\odot}$  ($4\times   10^{10}$
$M_{\odot}$  and  $3\times   10^{11}$  $M_{\odot}$).  Particles   were
initially displaced according to  the Zel'dovich approximation up to a
starting redshift $1+z_i=72$  ($51$,  $36$) for the $256^3$  ($128^3$,
$64^3$) grid. The initial gas density and velocity field was perturbed
according to the linear theory of  gravitational clustering, using the
same density and  displacement fields as for  dark matter. The initial
gas   temperature was set   to $T =  548(1+z_i)^{-2}$  K,  in order to
recover  the correct  thermal history of  baryons after recombination,
when   neglecting  re-ionization.  The  adiabatic   index  was  set to
$\gamma=5/3$, and a  fully ionized,  primordial $\rm  H$ and $\rm  He$
plasma was considered, with mean molecular weight $\mu = 0.59$.

\begin{table}
\caption{
RAMSES parameters for our six $\Lambda$CDM simulations.  In each case,
the box size was set to $L=100$ h$^{-1}$ Mpc.}
\label{tableruns}
\begin{tabular}{@{}llclllllll@{}}
\hline
Name & $N_{\rm part}$          & $\ell_{\rm max}$ & $N_{\rm cell}$
& $\Delta x_{\rm min}$ & $N_{\rm step}$\\

 & & & & h$^{-1}$kpc & $\ell=0$ & $\ell_{\rm max}$ \\
\hline
P064L0  &  $64^3$ & 0 & $2.6\times 10^5$ & 1562. &  51 & 51 \\

P128L0 & $128^3$ & 0 & $2.1\times 10^6$ & 781.3 & 107 & 107 \\

P256L0 & $256^3$ & 0 & $1.7\times 10^7$ & 390.6 & 243 & 243 \\

P064L3  &  $64^3$ & 3 & $5.6\times 10^5$ & 195.3 &  67 & 493 \\

P128L4 & $128^3$ & 4 & $5.0\times 10^6$ & 48.82 & 148 & 2259 \\

P256L5 & $256^3$ & 5 & $4.1\times 10^7$ & 12.21 & 304 & 7281 \\
\hline
\end{tabular}
\end{table}

\subsection{Refinement Strategy}

The main ingredient in the cosmological  simulations presented here is
the refinement strategy.  In order to  increase the spatial resolution
within   collapsing  regions,  a   quasi-Lagrangian mesh  evolution was
naturally chosen, using  level  dependent dark matter and  gas density
thresholds, as  explained  in  Section~\ref{sectrefinecriteria}. To be
more specific,  the level  dependent  density thresholds for the  dark
matter component were set to
\begin{equation}
\rho_{\ell} = 40 \frac{\Omega_m-\Omega_b}{\Omega_m} \frac{M_0}{(\Delta x^{\ell})^3}
\end{equation}
while for the baryonic component, they were set to
\begin{equation}
\rho_{\ell} = 40 \frac{\Omega_b}{\Omega_m} \frac{M_0}{(\Delta x^{\ell})^3}
\end{equation}
In this way, a roughly  constant number of particles per cell (between
5  and 40) is  obtained, minimizing  both collisionality  and particle
discreteness effects.  Note  that this number is higher  than the pure
dark  matter  simulations  performed  in \citet{kravtsov97},  where  5
particles were used to trigger  new refinements, instead of 40 in this
paper.  As explained above, this  choice has basically two reasons: 1-
we  prefer to  minimize as  much as  possible the  effect  of particle
discreteness effect  (Poisson noise) on  fluid dynamics, 2-  since the
memory storage  is dominated by  fluid variables, we can  increase the
number of  particles per cell  by one order  of magnitude for  a given
spatial resolution.  The number  40 was finally retained to allow
for a simple comparison with the more standard refinement threshold of
\citet{kravtsov97},  namely   a  factor  of  2   decrease  in  spatial
resolution.

Shock  refinements, as  discussed above,  was not  retained  for these
cosmological simulations.   This choice  has two reasons.   First, the
memory overhead  associated to shock refinements would  have been very
large,  since  shock  fronts  occur  everywhere  in  the  hierarchical
clustering picture. Since shock front are essentially two-dimensional,
the number  of cells required to  cover the shock  surfaces would have
been completely  out of reach,  even for modern  computers.  Secondly,
refining the mesh in low  density regions where shock waves eventually
propagate  would  violate the  non-collisionality  condition for  dark
matter dynamics. On  the other hand, the AMR  dynamics of shock fronts
in this case (no shock  refinements) was carefully investigated in the
last sections.  It turned out that as soon as shock waves travels from
high-resolution  to   low-resolution  regions,  no   spurious  effects
occurs. This last conditions turns out to be satisfied in cosmological
simulations, as discussed in the next section.

The   3 different initial   particle   grids considered here  ($64^3$,
$128^3$ and  $256^3$) defines also the coarse  level ($\ell=0$) of the
AMR hierarchy in each run. The maximum level  of refinement was set to
3, 4  and 5 respectively. This corresponds  to  a formal resolution of
$512^3$, $2048^3$ and $8192^3$ in the  highest resolution regions. The
corresponding  spatial resolution is 195, 48  and 12 h$^{-1}$ kpc.  In
all cases, adaptive time  stepping was  activated, with the  following
time step control parameters  $C_1=C_2=cfl=0.5$.  In order to  measure
the  advantage of adaptive mesh  in  cosmological simulations, these 3
runs were also performed without refinement ($\ell_{\rm max}=0$).

\subsection{Energy Conservation and Adaptive Mesh Evolution}
\label{econs}

\begin{figure}
\caption{
Top: Energy  conservation as  a function  of expansion  factor for run
P256L5   (solid  line),  P128L4  (dotted   line)   and P064L3  (dashed
line).  Bottom:  Total  number  of  cells in  the   AMR grid hierarchy
(including split cells) divided by the initial  number of coarse cells
as a function of expansion factor for the same 3 runs.  }
\label{econsandncell}
\end{figure}

A  standard measure  of the quality  of a  numerical simulation  is to
check for total energy conservation errors.  Since Euler equations are
solved  in  conservative  form in RAMSES,   the main  source of energy
conservation errors comes   from the gravitational  source terms,  for
both baryons  and dark  matter components.  Figure~\ref{econsandncell}
shows the total energy conservation  (in the form of the Layzer-Irvine
conservation  equation  for an  expanding   universe)  for the  3  AMR
runs. The maximum errors  are found to   be 2~\%, 1~\% and  0.5~\% for
P064L3,  P128L4  and P256L5  respectively.   The maximum  error in the
energy conservation occurs  when  a significant number of  refinements
are built for  the  first time, around  $1+z  \simeq$ 2, 3 and  5  for
P064L3, P128L4 and P256L5 respectively.

In Figure~\ref{econsandncell} is also shown  the total number of cells
in the  AMR tree structure  (including  split cells), in units  of the
number of coarse cells. It is  worth noticing that this numbers should
have remained exactly equal to 1 for a strict Lagrangian mesh like the
ones  described in \citet{gnedin95}  and \citet{pen95}.  At the end of
the  simulations, as clustering  develops,  the final  number  of mesh
points has increased by a factor of 2.5.  This  overhead is related to
the mesh smoothing operator (see Sect.~\ref{sectrefinemap}).  The mesh
evolution is therefore only ``quasi'' Lagrangian.

\subsection{Adaptive Mesh Structure}

\begin{figure*}
\centering
\caption{
Gray  scale images of  the  gas temperature ($T   > 10^5$ K), the  gas
density ($\rho > \Omega_b \rho_c$)  and the dark matter density ($\rho
> \Omega_m \rho_c$) in a  planar cut through the  computational volume
for run P256L5. The mesh structure within  the plane is plotted in the
upper    right  panel  (only    octs  boundaries    are  shown  for
visibility). For sake of comparison,  the density and temperature maps
obtained for run  P256L0 (same initial conditions without refinements)
are shown in the 2 lower panels.  }
\label{figlargescale}
\end{figure*}

The adaptive mesh is dynamically modified at each time step during the
course of the simulation.  Both hydrodynamics and N-body  solvers take
advantage of the increased spatial  resolution to improve the accuracy
of the solution in the refined regions.

Figure~\ref{figlargescale} illustrates this by  comparing the gas  and
dark matter    density   fields  in  a    slice  cutting   through the
computational volume for run P256L5  ($256^3$ particles with 5  levels
of refinements)  with  the density fields  in the  same  slice for run
P256L0 (same initial conditions  without refinements). Only  overdense
cells  are shown ($\rho   > \Omega_b \rho_c$  for  baryons and $\rho >
(\Omega_m - \Omega_b) \rho_c$ for dark matter).  One clearly sees that
both gas and dark matter density fields are much more dense and clumpy
when refinements are activated. On the other hand, it is reassuring to
see that both simulations agree with each other on large scale.

The corresponding mesh structure  is shown in the  upper right part of
Figure~\ref{figlargescale}.    The  interest of   using a   tree-based
approach for  defining the   AMR   hierarchy is striking:  the    grid
structure closely  follows the geometry  of the  density field, from a
typical filamentary  shape  at large  scale, to a  more  spherical and
compact  shape in the higher density  haloes cores. If one examines the
central   filament  connecting the 2 massive   haloes  in the images of
Figure~\ref{figlargescale}, one  sees    that it   follows a   typical
pancake-like structure, with 2 dark matter caustics on  each side of a
gas filament. This structure, though  interesting, is not dense enough
to be refined  by our refinement  strategy.  We could have lowered the
density thresholds to  trigger new refinements in  this region, at the
price of an  increased   collisionality for dark matter.    This would
result in a spurious fragmentation of the pancake structure
\citep{melott97}.

The temperature map ($T > 10^5$ K) in the same planar cut exhibits the
typical flower-like structure of  strong cosmological accretion shocks
around large haloes. These    strong shocks propagates  exclusively  in
large voids  in between  filaments  that intersect each  other  at the
central halo position: this  is due to  the higher gas pressure within
the filaments that inhibits shock propagation in the direction aligned
with  the filaments. This property of   cosmological shock waves is of
great importance here, since it  implies that strong shocks propagates
almost exclusively from high to low resolution regions of the grid. On
the other hand,  weak shocks occurring  during sub-halo mergers along
the  filaments  can  enter high  resolution  regions  of  the  mesh as
clustering develops.  Since   the oscillatory   behavior  outlined  in
Section~\ref{shockcoarsetofine} disappears completely for weak shocks
\citep{berger89,khokhlov98},  we can conclude that cosmological shocks
propagation remains  free from   spurious  effects associated  to  the
adaptive mesh structure.

\subsection{The Halo Model}
\label{sec:halomodel}

In    order to analyze   the  results  of the    simulations in a more
quantitative  way,  I will  use  a powerful analytical  theory: the so
called {\em  halo   model}.   Several authors  \citep{seljak00,  ma00,
scoccimarro01, cooray00, cooray01, refregier01} have recently explored
the idea  that  both dark   matter  and baryons  distributions can  be
described  by the  sum  of  two contributions:  (1)   a collection  of
virialized, hydrostatic haloes with overdensity $\ge 200$ and described
by  the Press \& Schechter  mass function and  (2) a smooth background
with   overdensity  $\le 10$  described   by   the linear   theory  of
gravitational clustering.  The purpose of this paper is not to improve
upon earlier works  on this  halo model, but  rather to  use it as  an
analyzing tool for our  simulations. Therefore, the  basic ingredients
of the halo model are only briefly recalled  and will not be discussed
in  great details. From now on,  we consider  only results obtained at
the final redshift $z=0$. The redshift  evolution of the halo model is
discussed  and      compared   to  numerical   simulations   elsewhere
\citep{refregier00,refregier01}.

Haloes are defined  as virialized clump  of gas  and dark matter,  with
total mass defined as the Virial mass
\begin{equation}
M_{\rm vir} = \frac{4\pi}{3} \Delta_c \rho_c r_{\rm vir}^3
\end{equation}
where the Virial  {\em density contrast}  $\Delta_c$ is related to the
Virial {\em  overdensity}   by  $\Delta_c =   \Omega_m  \Delta$.   For
$\Omega_m =  0.3$     and $z=0$,    one has     $\Delta  \simeq   334$
\citep{eke98}. The  dark matter follows  the  Navarro, Frenk \&  White
\citep[NFW;][]{navarro96} density profile
\begin{equation}
\rho = \frac{\rho_s}{ (r/r_s)(1+r/r_s)^2 }
\end{equation}
whose parameters are connected to the halo Virial mass by
\begin{equation}
M_{\rm vir} = 4\pi \rho_s r_s^3 \left[ \ln (1+c) - \frac{c}{1+c}
\right]
\end{equation}
The only remaining free parameter (apart from $M_{\rm vir}$) is the so
called concentration parameter $c = r_{\rm vir} / r_s$. This parameter
exhibits a  good correlation with  halo  mass in numerical simulations
that  can  be fitted   analytically  to  match the  numerical  results
\citep{ma00}.   The   final ingredient  is   to  assume that  the halo
distribution  is described  by  the Press  \& Schechter  mass function
\citep{press74}.

The total  mass power  spectrum can  then  computed  as the sum   of 2
components  $P(k) = P_1(k)  +P_2(k)$,  where $P_1(k)$ is a  non-linear
term corresponding to the mass  correlation within halos, and $P_2(k)$
is a  linear  term corresponding  to  the  mass  correlation between  2
halos.    Both   terms have    relatively  straightforward  analytical
expressions  that  are    not recalled  here   \citep{seljak00,  ma00,
scoccimarro01}.

The gas distribution within halos is supposed to follow the isothermal
hydrostatic  equilibrium.  The temperature remains constant  within the
halo Virial radius, and is taken equal to the halo Virial temperature
\begin{equation}
\frac{k_B T_{\rm vir}}{\mu m_H} 
= \frac{1}{2} \frac{GM_{\rm vir}} {r_{\rm vir}}
\label{virtemp}
\end{equation}
Note that temperature profiles determined in numerical simulations are
neither  isothermal nor equal to the   Virial temperature.  Therefore,
the   halo  model can  only  be  considered as  a crude approximation,
describing the gas distribution in a statistical sense only. Moreover,
the temperature  profiles observed in  large X-ray clusters is more or
less affected by cooling flows  in the central regions. Including  all
physical ingredients that  might affect the  thermal structure in  the
core  of virialized  halos is  beyond the  scope  of  this paper. Only
adiabatic hydrodynamics is considered here.

Solving  the hydrostatic equilibrium  equation in  the isothermal case
(using the NFW mass distribution)  leads to the following gas  density
profile \citep{suto98}
\begin{equation}
\rho = \rho_0 {\rm e}^{-b} \left( 1+ r/r_s \right)^{b r_s /r}
\label{anagasprof}
\end{equation}
where the dimensionless parameter $b$ is given by
\begin{equation}
b = \frac{\mu m_H 4\pi G \rho_s r_s^2}{ k_B T_{\rm vir}}
\end{equation}
The central density $\rho_0$ is computed  by specifying that the total
baryons mass within the Virial radius is equal to $\Omega_b / \Omega_m
M_{\rm vir}$.

In the next section, the gas  pressure power spectrum is computed from
RAMSES  numerical   simulations   and  compared  to  the   halo  model
predictions. The  pressure   power spectrum  is  quite an  interesting
quantity  in    cosmology,  since it    is   directly related   to the
Sunyaev-Zeldovich  induced Cosmic Microwave Background    anisotropies
angular power spectrum  \citep{sunyaev80, rephaeli95}. It can be computed within
the halo model framework  using the same two  terms  as for  the total
mass density power spectrum $P(k) =  P_1(k) +P_2(k)$. Exact analytical
expressions can be found in \citet{cooray01} and \citet{refregier01}.

\subsection{Power Spectra}

\begin{figure*}
\centering
\caption{ 
Dark   matter density (left panels)   and gas  pressure (right panels)
power spectra for RAMSES   runs  with refinements (upper  panels)  and
without   refinements  (lower panels). In  each   plot, the solid line
corresponds    to the halo model   prediction,   while the dotted line
(dashed and dot-dashed) corresponds  to numerical results with $256^3$
($128^3$  and $64^3$) particles.    Label ``AMR'' stands  for AMR runs
(P256L5, P128L4 and P64L3), while label ``PM'' stands for runs without
refinement (P256L0,  P128L0 and P64L0).  For each power  spectrum, the
curve ends  at the   Nyquist frequency  corresponding to   the formal
resolution    of  the    simulation   ($\Delta     x_{\rm   min}$   in
Table~\ref{tableruns}).  }
\label{figpowerspectra}
\end{figure*}

In this   section, both dark  matter density   and gas  pressure power
spectra are  computed and compared to  the  halo model predictions. In
order to  study the convergence properties  of the numerical solution,
results obtained   with different  mass  and  spatial  resolutions are
examined.

\subsubsection{Computing the power spectra}

Computing power spectra for simulations with such high dynamical range
requires to go beyond traditional  methods based on regular  Cartesian
meshes: recall that our highest resolution run has a formal resolution
of $8192^3$. We use instead a multi-grid method based on a hierarchy of
nested cubic  Cartesian grids  \citep{jenkins98,  kravtsov99}.  Each
level of  the  hierarchy corresponds  to  the code AMR   levels (from
$\ell=0$    to $\ell_{\rm max}$) and   covers  the whole computational
volume with $\ell^3$ cubic grids of size $n_x^3$.  At a given level,
a  dark matter  density  field is  computed  for  each grid  using CIC
interpolation,   and   all grids  are stacked   together  in a single,
co-added  density field. This density  field  is then Fourier analyzed
using  FFT technique.  From  the  resulting  power spectrum, only  modes
spanning  the  range $2^\ell  \times \left[  k_{\rm  min}, k_{\rm max}
\right]$ are kept as reliable estimations of  the true power spectrum,
with $k_{\rm    min} = k_{\rm nynq}/8$   and  $k_{\rm  max}   = k_{\rm
nynq}/4$.  The  Nyquist frequency $k_{\rm nynq}$  depends on the size
of the cubic  grid,  chosen  here equal to   the  coarse grid  size
$n_x^3$, so that $k_{\rm   nynq}=n_x\pi /L$. At the   2 extreme
spatial  scales,  we have  however  $k_{\rm  min}  =  2\pi /  L$  (for
$\ell=0$)  and  $k_{\rm max}  =  2^{\ell_{\rm max}} k_{\rm nynq}$ (for
$\ell=\ell_{\rm  max}$).  The    maximum frequency  considered  in the
present  analysis  corresponds    therefore to  the   formal  Nyquist
frequency of each simulation $k_{\rm max}  = \pi / \Delta x_{\rm min}$
(see  Table~\ref{tableruns}).  The  same procedure  is applied to  the
pressure field, except that CIC interpolation is no longer needed.

\subsubsection{Results}
\label{powerresults}

The resulting power spectra  are shown in Figure~\ref{figpowerspectra}
for the   3 runs with    refinements  (labeled ``AMR'')   and  without
refinements (labeled ``PM'').    For comparison, the  dark  matter and
pressure power spectra   predicted by the  halo  model are  plotted as
solid lines.

The dark matter power  spectrum  shows a  striking agreement with  the
halo model prediction, down to the  formal resolution limit. Note that
the  halo model  free  parameters has been tuned  in  order to recover
simulations results \citep{ma00}.  Results obtained here are therefore
consistent   with those obtained   by  other authors \citep{jenkins98,
kravtsov99}, and  can be considered as a  powerful integrated test of
the code.  For each mass resolution, the  power spectrum is plotted up
to the  formal Nyquist frequency.    For  AMR runs with   refinements
activated, the numerical   power  spectrum is dominated at   high wave
numbers by the Poisson noise due to particle discreteness effects (see
the small  increase of power around  $k_{\rm nynq}$).  We can conclude
that the numerical power spectrum has converged for  each run, down to
the limit imposed   by  the  finite  mass resolution,   without  being
affected by  the    finite   spatial resolution. For    runs   without
refinements, the limited  dynamical range has  a noticeable  effect on
the resulting  power spectrum   at   much larger scale:   the  spatial
resolution is therefore a strong limiting factor in this case.

Let   us   now     examine    the   gas pressure     power    spectrum
(Fig.~\ref{figpowerspectra}). The  overall agreement of  the numerical
results  with  the  halo  model predictions  is  relatively  good: the
correct  behavior is captured at all  scales within a  factor of 2 for
our highest resolution run  (P256L5). Note that  the halo model has no
free parameters for  the gas distribution, as soon  as the dark matter
parameters are  held fixed.   At   large scale, the   numerical  power
spectrum appear to converge to  the halo model predictions (run P128L4
and run P256L5 give exactly the same results). Note that a rather high
mass resolution  is needed for   this convergence to occur  ($> 128^3$
particles),  as opposed to the dark  matter density power spectrum for
which the correct   large scale power is   recovered even with  $64^3$
particles.  At intermediate scales  (around  1 h Mpc$^{-1}$), the halo
model predictions slightly underestimate  the pressure power spectrum.
Since  numerical  results have also  converged  at  these scales, this
discrepancy might be due to the fact that intermediate density regions
($10 < \rho < 200$) are completely  neglected in the halo model. These
regions are believed to be  composed of warm filaments, whose pressure
obviously cannot be completely neglected.

At small scales, the situation remains quite unclear. On one hand, one
clearly  sees   on  Figure~\ref{figpowerspectra}  that   an  increased
dynamical range has a dramatic  effect on the resulting pressure power
spectra.  The power  is much higher on small scales  for AMR runs than
for runs  without refinements.  On the  other hand, for  AMR runs, the
convergence of the  numerical results to the ``true''  solution is not
as  fast as  that of  the dark  matter power  spectrum. Some  hints of
convergence  between  run  P128L4  and  run  P256L5  can  be  seen  on
Figure~\ref{figpowerspectra}. Indeed, without refinements (runs P64L0,
P128L0  and P256L0), the  cut off  in the  pressure power  spectrum is
directly proportional to the spatial resolution of the simulation. The
same is true between runs P64L3  and P128L4, while for run P256L5, the
effect of  spatial resolution appears  to weaken slightly.  More
interesting is the large discrepancy in slope at large $k$ between the
halo model and  the solution obtained by run  P256L5.  As discussed in
the  next  section,  this  is   probably  due  to  the  assumption  of
isothermality  in the  halo model,  which is  ruled out  by simulation
results for individual halo temperature profiles. 

\subsection{Individual Halos Structure}

\begin{figure*}
\vspace{1cm}
\centering
\vspace{1cm}
\caption{
Color maps showing various projected quantities for the 5 most massive
halos extracted from run P256L5. The  projected volume is in each case
a cube  of  6.25 h$^{-1}$Mpc aside. The  color  coding is based  on  a
logarithmic scale for each plot. From top to bottom: 1- Projected dark
matter  particle distribution  (the  color coding  corresponds  to the
local particle density). 2- X-ray  emissivity map. 3- X-ray  (emission
weighted)  temperature  map. 4-  Sunyaev-Zeldovich decrement parameter
(equivalent  to the integrated pressure  along  the line of sight). 5-
Projected adaptive mesh structure (only oct boundaries are shown).  }
\label{fighalosmaps}
\end{figure*}

\begin{figure*}
\vspace{1cm} 
\centering
\vspace{1cm}
\caption{
Radial   profiles   for the  5 most    massive  halos  showing various
quantities averaged along  radial bins. In each  plot, the dotted line
(dashed and dot-dashed) comes from run P256L5 (P128L4 and P64L3). From
top to bottom: 1- Dark   matter overdensity profile (the best-fit  NFW
analytical  profile  is shown  as a  solid   line). 2- Gas overdensity
profile (the corresponding ``hydrostatic isothermal model'' analytical
profile is also shown as  a solid line).  3- Mass averaged temperature
profile  (the corresponding   ``hydrostatic  beta model''   analytical
profile is  also  shown as a  solid  line). 4-   Pressure profile (the
corresponding ``hydrostatic isothermal  model'' analytical profile  is
also shown as a solid line). 5- Volume averaged refinement levels.  }
\label{fighalosprofiles}
\end{figure*}

The internal  structure of the  5  largest halos  found in the highest
resolution run (P256L5) is now examined  in great detail. It is worth
mentioning that this   analysis is made  possible  thanks to the large
dynamical range obtained in our simulation  ($\Delta x_{\rm min} =$ 12
h$^{-1}$ kpc), although it was not  optimized to study individual halo
properties. The next step to go beyond what is presented here would be
to  perform  so   called ``zoom  simulations'',   with  nested, higher
resolution, initial conditions particle grids centered on single halos
\citep{bryan97,eke98,yoshida00,abel00}.   The  main  advantage of  the
present  ``brute force'' approach is to  combine  both large and small
scale   results in the analysis.   In  order to  study the convergence
properties  of individual halo   profiles, results  obtained for  runs
P256L5,  P128L4  and  P64L3   are compared.    Recall that   a  direct
comparison of the same  halo at different  mass and spatial resolution
is possible, since the same initial conditions were used (and degraded
to the correct mass resolution) for each run.  Runs without refinement
(P256L0, P128L0 and  P64L0) are discarded  from this analysis, because
they  completely lack  the necessary  dynamical range  to resolve  the
internal structure of individual halos.

\begin{table}
\caption{
Global properties of  the 5 largest  halos extracted from the  highest
resolution run P256L5.}
\label{tablehalos}
\begin{tabular}{@{}llllllllll@{}}
\hline
Name & $M_{\rm vir}$ & $r_{\rm vir}$ & $c$ & $\beta_{\rm fit}$ &
$r_{\rm core}$ \\
 & h$^{-1}$M$_{\odot}$ & h$^{-1}$Mpc &  &  & h$^{-1}$kpc \\
\hline
Cluster 1 & $6.97\times 10^{14}$ & 1.82 & 9.5 & 0.85 & 144.2 \\

Cluster 2 & $5.09\times 10^{14}$ & 1.64 & 5.9 & 0.80 & 215.5 \\

Cluster 3 & $5.07\times 10^{14}$ & 1.63 & 7.3 & 0.80 & 202.3 \\

Cluster 4 & $4.52\times 10^{14}$ & 1.57 & 4.9 & 0.64 & 147.0 \\

Cluster 5 & $4.29\times 10^{14}$ & 1.55 & 9.4 & 0.84 & 159.7 \\

\hline
\end{tabular}
\end{table}

Haloes were detected in the dark matter  particles distribution at the
final output time  ($z=0$)  using  the Spherical Overdensity algorithm
\citep{lacey93}, with overdensity threshold $\Delta = 334$. Only the 5
most massive  haloes of the resulting mass  function are considered for
the present    analysis.   Their global   properties    are listed  on
Table~\ref{tablehalos}.   For each halo, the  center is defined as the
location  of  the  maximum in the   dark  matter  density  field.  For
regular, relaxed haloes,   this   definition also  corresponds  to  the
maximum in the baryons density field, and to the  halo center of mass.
This is however not the case for irregular, not yet relaxed haloes, for
which this  definition of the  halo   center is  less  robust.   

Cubic regions 6.25  h$^{-1}$ Mpc aside  are then extracted around each
halo center.  The resulting  projected color maps for various relevant
quantities are  shown in Figure~\ref{fighalosmaps}.   Clusters 1 and 5
appear to be the most relaxed halos of our sample, while clusters 2, 3
and 4 show more substructures  and irregularities within their  Virial
radii.    The   adaptive   mesh     structure,     also   shown     in
Figure~\ref{fighalosmaps},  closely  matches the  clumpy  structure of
each halo.   Note that the  maximum level of  refinement ($\ell=5$) is
activated  in  the  halo cores, where    the formal spatial resolution
reaches 12 h$^{-1}$ kpc  (barely visible in  Fig.~\ref{fighalosmaps}).
The physical     properties of  these  5   haloes    are now  discussed
quantitatively.

\subsubsection{Dark Matter Distribution}

The projected dark    matter particles   distribution is  shown     in
Figure~\ref{fighalosmaps},  with  a color coding  corresponding to the
local particle overdensity.  The dark matter  distribution is far from
being smooth and spherically symmetric.   It is however interesting  to
compute  the    radial density  profile and   compare   it  to the NFW
analytical      prediction.       The   result       is   shown     in
Figure~\ref{fighalosprofiles}, for  our 3 different  mass  resolutions
(runs P256L5, P128L4 and P64L3).  The  density profile obtained in the
highest resolution  run (P256L5) was  fitted to the NFW profile, using
the concentration parameter $c$ as fitting  parameter. Best fit values
are listed in  Table~\ref{tablehalos}: they are fully  consistent with
expected values for a $\Lambda$CDM universe \citep{kravtsov97, eke98}.

The quality of the fit  is impressive for clusters  1 and 5, which are
also  the   more relaxed haloes   of  our sample.  The  poorest fit was
obtained for cluster 4, with deviations as large  as 50~\% at a radius
of 150 h$^{-1}$ kpc.  A close examination of  the corresponding map in
Figure~\ref{fighalosmaps} confirms that this halo is poorly relaxed in
its central region. By comparing  the  profiles obtained for  different
mass resolution, one sees that the  numerical profiles agree with each
other down to  their resolution limit. This  is in complete  agreement
with our    conclusion  concerning the  dark  matter   power spectrum:
simulated power spectra match closely the halo model prediction down
to their formal Nyquist frequency.

\subsubsection{Baryons Distributions}

The baryons distribution within the selected haloes is similar on large
scales to the dark matter distribution.  In Figure~\ref{fighalosmaps},
simulated X-ray  emission maps (using   $L_X \propto n_e^2$) are  good
tracers of the gas overdensity projected  along the line of sight. One
can notice however that the hot gas is  more smoothly distributed than
dark matter.  This is even more striking in  the central region of all
clusters,  where the gas density reaches  a plateau,  reminiscent of a
$\beta$ model density profile. It is worth noticing that overdensities
(substructures) in  the gas distribution usually  appears as cold spots
in the X-ray (emission  weighted) temperature map.  On the other hand,
the cluster cores are significantly hotter than the surrounding gas in
most cases.  We will come back to this point later.

Using the halo center as defined above, gas overdensity, mass weighted
temperature and  pressure profiles were   computed  as a  function  of
radius    and plotted   in  Figure~\ref{fighalosprofiles}.  For   that
purpose,  conserved variables such   as mass and  internal energy  are
averaged into radial bins  of increasing thickness, starting from  the
formal resolution (12  h$^{-1}$ kpc for run   P256L5) up 3 times  this
value at the Virial radius.   The volume averaged refinement level  is
also plotted  in Figure~\ref{fighalosprofiles}, giving  some hints  of
the effective spatial resolution as a function of radius. Based on the
results obtained during the Spherical  Secondary Infall test presented
in   \S~\ref{sectsecinfall},   the actual   resolution    of the  code
corresponds roughly to twice  its  formal resolution.  For  run P256L5
(P128L4 and P64L3),  this gives a limiting  radius of 24 (96  and 384)
h$^{-1}$ kpc, above which numerical results are fully reliable. 

Since dark matter density profile are well fitted for each halo by the
NFW analytical  profile, the corresponding gas density  profile can be
computed using  Equation~(\ref{anagasprof}).  Recall that  in the halo
model, the  gas temperature is  assumed to remain constant  within the
Virial    radius,    and    equal    to   the    Virial    temperature
(Eq.~\ref{virtemp}). This is obviously  not the case for our simulated
clusters (see Fig.~\ref{fighalosprofiles}),  and explains why the halo
model  profile is  much more  peaked in  the central  region  than for
simulated  profiles.   For the  pressure  profiles,  the situation  is
however less dramatic, though  still unsatisfactory. It is interesting
to notice  that here again, the  behavior of the  pressure profiles is
fully consistent  with previous results concerning  the pressure power
spectrum: the  isothermal halo model overestimates  the pressure power
on small scales.  This translates in a much  steeper slope at low
radii, as  for the pressure  power spectrum at large  $k$.  Moreover,
numerical results for the  gas distribution within haloes have clearly
not converged  yet, although the  dependance of the  computed profiles
with  respect  to the  spatial  resolution  seems  to weaken  slightly
between run  P128L4 and  run P256L5,  in exactly the  same way  as the
pressure power spectrum did in \S~\ref{powerresults}.  
Conclusions  are therefore similar: numerical
results show  good evidence  of converging at  scales greater  than 50
h$^{-1}$  kpc.  Similar conclusions  were obtained  by \citet{bryan97}
for a ``zoom'' simulation of a single cluster.  These authors obtained
very similar gas density and temperature profiles, with quite the same
convergence properties  as the  one obtained here.   Consequently, the
isothermal  halo model  is  likely to  fail  at scales  less than  250
h$^{-1}$ kpc.

\subsubsection{Beyond the Isothermal Halo Model ?}

As   noted by several authors  \citep{bryan97,eke98},  the typical gas
density profile  obtained    in numerical simulations   is  much  more
accurately    described by  the  $\beta$    model analytical  solution
\citep{cavaliere76}
\begin{equation}
\rho = \rho_0 \left[ 1 + (r/r_{\rm core})^2 \right]
^{-3\beta_{\rm fit}/2}
\label{anagasbeta}
\end{equation}
Although the numerical results presented here have not fully converged
yet, it is  worth  exploring an  alternative  to the  isothermal  halo
model.  The  gas overdensity profile obtained  in run  P256L5 was thus
fitted with  the $\beta$ model formula,  using both $r_{\rm core}$ and
$\beta_{\rm  fit}$ as fitting parameters.   Best fit values are listed
in Table~\ref{tablehalos} and  are   consistent with  typical  numbers
quoted by other authors \citep[e.g.][]{eke98}.  It is worth mentioning
that the  quality  of the  fit is excellent  in each  case, except for
cluster 4,  as expected from the  previous analysis on the dark matter
distribution.

Since both gas and dark matter  density profiles are now determined to
a good accuracy,  it  is possible  to perform a  consistency check and
compute the temperature profile   resulting  from the  assumption   of
hydrostatic equilibrium.   For that  purpose, the analytical framework
developed  by  \citet{varnieres01} is  exactly  what is  needed here:
assuming  a NFW   profile for dark   matter  and a   $\beta$ model for
baryons, they  derived    analytically the corresponding   hydrostatic
temperature   profile.  This temperature   profile  is shown for  each
cluster in Figure~\ref{fighalosprofiles}. The agreement with numerical
results  is good: temperature rising   towards the halo center  is
therefore a  direct   consequence of hydrostatic    equilibrium. After
closer examination, the small (20~\%) disagreement observed in clusters
2, 3  and 4 is due  to the poorer fit of  the NFW  formula to the dark
matter simulated density profiles.

This non  constant behavior of  the halo temperature profile have been
already   noticed   by  several    authors  in   numerical simulations
\citep{bryan97,eke98, loken00} and also in X-ray observations of large
galaxy clusters \citep{markevitch98}.   Note however that more physics
need to be included in current numerical simulations before performing
a   reliable comparison  to  observations.    On the  other hand,  the
idealized case of adiabatic gas dynamics is still of great theoretical
interest: one can  hope to find  a self-consistent description of  the
gas  distribution within  haloes  using  such an approach.   The  main
ingredient to extend the halo model  in this framework is to determine
the typical core  radius (as well  as $\beta_{\rm fit}$) as a function
of halo mass and redshift.  \citet{eke98}  have already initiated this
challenging task   using high resolution  (zoom)   SPH simulations for
large mass haloes ($M \simeq 10^{15}  h^{-1}M_{\odot}$), but more work
need to be done to probe a larger mass and  redshift range.  Using the
analytical formula of \citet{varnieres01}, it would be straightforward
to determine  the corresponding   temperature   profile, and thus   to
complete the description of the baryons component within this extended
halo model.  This  ambitious project will not  be  addressed here, but
will be considered in a near future.

\section{Conclusions and Future Projects}
\label{sec:conclusion}

A  new  N-body/hydrodynamical code,  RAMSES,  has  been presented  and
tested  in various  configurations. RAMSES has  been  written in
FORTRAN 90 and optimized on  a vectorized hardware, namely the Fujitsu
VPP 5000 at  CEA Grenoble.  A parallel version  was also impemented on
shared-memory  systems using OpenMP  directives.  A  distibuted memory
version  of RAMSES  is  currently under  construction  using a  domain
decomposition approach.

The main features of the RAMSES code are  the followings: 

1- the AMR  grid is built on  a  tree structure, with new  refinements
dynamically  created  (or  destroyed) on a   cell-by-cell  basis. This
allows greater flexibility to match complicated flow geometries.  This
property    appears   to  be   especially   relevant   to cosmological
simulations,  since clumpy structures    form and collapse  everywhere
within the hierarchical clustering scenario.

2-  the hydrodynamical  solver  is  based on  a  second order  Godunov
method,  a modern  shock  capturing method  that  ensures exact  total
energy  conservation, as  soon  as  gravity  is  not  included.
Moreover,  shock capturing  relies on  a Riemann  solver,  without any
artificial viscosity.

3-  the refinement   strategy   that was   retained   for cosmological
simulations is based on a ``quasi-Lagrangian''  mesh evolution. In this
way, the  number of  dark  matter particles  per cell remains  roughly
constant, minimizing two-body  relaxation and  Poisson noise.  On  the
other hand, this refinement strategy is not optimal for baryons, since
one neglects to refine  shock fronts (this would  have been too costly
anyway).   It has been  carefully shown that in  this case, as soon as
strong  shocks propagate  from high  to low resolution  regions of the
grid, no spurious effects appear.

The code  has been tested in  standard  gas dynamical test cases (Sod's
test and Sedov's test),  but  also for integrated cosmological  tests,
like Zel'dovich pancake collapse  or Bertschinger  spherical secondary
infall.  It  has been shown  that  the actual resolution limit  of the
code is equal to roughly twice the cell size of the maximum refinement
level.

The RAMSES code   has been finally  used   to study  the  formation of
structures     in a low-density      $\Lambda$CDM universe.  A careful
convergence   analysis has been   performed,   using the same  initial
conditions with various mass and spatial resolutions,  for a fixed box
size $L=100$ h$^{-1}$ Mpc.  The initial number of  cell (at the coarse
level) was set equal to the initial  particle grid ($64^3$, $128^3$ or
$256^3$), for  a final number  of cells only  2.5  larger.  The formal
spatial resolution in the largest run was $8192^3$  or 12 h$^{-1}$ kpc
comoving.

Numerical results have been  compared to the analytical predictions of
the so called halo model, for  both dark matter and gas pressure power
spectra,  as well  as  individual haloes  internal  structure. A  good
agreement was found  between the halo model and  the numerical results
for dark matter, down to  the formal resolution limit. For the baryons
distribution, numerical  results show  some evidence of  converging at
scales greater  than 50 h$^{-1}$  kpc for our highest  resolution run.
The  halo model  reproduces simulations  results  only approximatively
(within a factor of 2) at these scales.

A simple extension of the halo model  for the fluid component has been
proposed. The idea is to  assume that the  average gas density profile
within haloes is described by a $\beta$  model, whose parameters still
need  to   be  determined from   first  principles  or  from numerical
simulations and for a rather large mass range, which is far beyond the
scope of this paper \citep[see however][]{eke98}.  It is then possible
to  deduce  from hydrostatic  equilibrium   an analytical  temperature
profile  \citep{varnieres01}  that accurately  matches the  simulation
results presented    in this  paper,    and  should therefore  improve
considerably the halo model.

Extending the current work to ``zoom''  simulations is currently under
investigation, using  a set of nested grids  as initial conditions, in
order to improve  mass   and  spatial resolutions  inside   individual
haloes.   This   approach seems indeed very   natural   within the AMR
framework, and has already proven to  be successful in recent attempts
\citep{bryan97,abel00}.  Future efforts in the RAMSES code development
will  be  however  more   focused on including    more physics  in the
description of the gaseous component, like cooling, star formation and
supernovae feedback.

\begin{acknowledgements}
The simulations presented in this  paper were performed on the Fujitsu
VPP 5000 system of the Commissariat \`a l'Energie Atomique (Grenoble).
I would  like to thank Philippe  Kloos for his help  on optimizing the
code.   I would  like to  thank Remy  Abgrall for  inviting me  at the
CEMRACS Summer School  2000 in Luminy (Marseille), where  part of this
work  was initiated.  Special  thanks to  Alexandre Refregier  who has
provided me the software to compute analytical power spectra predicted
by the halo  model, and to St\'ephane Colombi who  has provided me the
software to  compute numerical power  spectra.  I would like  to thank
Jean-Michel Alimi, Edouard Audit,  Fran\c cois Bouchet and Jean-Pierre
Chi\`eze  for  enlightening discussions  about  the numerical  methods
discussed  in this  paper.  The  author is  grateful  to an  anonymous
referee, whose comments greatly improved the quality of the paper.
\end{acknowledgements}

\bibliographystyle{apj}
\bibliography{romain}

\label{lastpage}

\end{document}